\documentclass[12pt]{article}
\usepackage{amssymb}
\usepackage{a4}
\evensidemargin \oddsidemargin
\def\1{{\bf 1}}
\def\0{{\bf 0}}
\def\id{\mbox{id\,}}
\def\ot{\!\otimes\!}

\def\F{\mbox{$\cal F$}}

\def\cross{\mbox{$\rule{0.7pt}{1.3ex}\!\times $}}

\def\cross{\mbox{$\rule{0.5pt}{1.1ex}\!\times \,$}}
\def\ra{\rangle}

\def \A{\mathcal{A}}
\def \B {\mathcal{B}}
\def \C {\mathcal{C}}

\def \H {\mathcal{H}}

\def \I {\mathcal{I}}
\def \Oo {\mathcal{O}}

\def \L {\mathcal{L} }

\def \V {\mathcal{V} }

\def \X {\mathcal{X} }
\def \Y {\mathcal{Y}}

\def\Ha{{\sf H}}

\newcommand{\trc}{\triangleright}
\def\g{\mbox{\bf g\,}}
\def\gQ{\mbox{\bf g}_{\scriptscriptstyle Q}\,}
\def\GQ{G_{\scriptscriptstyle Q}\,}
\def\YQ{Y_{\scriptscriptstyle Q}\,}
\def \YYQ {\mathcal{Y}_{\scriptscriptstyle Q}\,}
\def \OQ {\mathcal{O}_{\scriptscriptstyle Q}\,}
\def\HQ{\H_{\scriptscriptstyle Q}}
\def\HeQ{H_{\scriptscriptstyle Q}}
\def\heQ{{\bf h}_{\scriptscriptstyle Q}}
\def\bA {\beta^{\scriptscriptstyle A}}
\def\bS{\beta^{\scriptscriptstyle S}}

\def\cp {{\stackrel{ {\stackrel{~}{\bullet}} }{+}}}

\newcommand{\bpsi}{\mbox{\boldmath $\psi$}}

\newcommand{\bbA}{\mbox{\boldmath $\beta$}^{\scriptscriptstyle A}}
\newcommand{\bnabla}{\mbox{\boldmath $\nabla$}}

\def\b#1{{\mathbb #1}}
\def\nn{\nonumber \\}

\newcommand{\be}{\begin{equation}}
\newcommand{\ee}{\end{equation}}
\newcommand{\bea}{\begin{eqnarray}}
\newcommand{\eea}{\end{eqnarray}}
\newcommand{\ba}{\begin{array}}
\newcommand{\ea}{\end{array}}
\newtheorem{prop}{Proposition}

\def\sq{\mbox{\rlap{$\sqcap$}$\sqcup$}}
\newenvironment{proof}[1]{\vspace{5pt}\noindent{\bf Proof #1}\hspace{6pt}}%
{\hfill\sq}
\newcommand{\bp}{\begin{proof}}
\newcommand{\ep}{\end{proof}\par\vspace{10pt}\noindent}
\begin{document}

\title{On quantum mechanics with a magnetic
field on $\mathbb{R}^n$ and on a torus $\mathbb{T}^n$, and their relation}

\author{   Gaetano Fiore,  \\\\
$^{1}$ Dip. di Matematica e Applicazioni, Universit\`a ``Federico II''\\
   V. Claudio 21, 80125 Napoli, Italy\\         
$^{2}$         I.N.F.N., Sez. di Napoli,
        Complesso MSA, V. Cintia, 80126 Napoli, Italy}
\date{}

\maketitle

\begin{abstract}
\noindent

We show in elementary terms the equivalence  in a general gauge of
a $U(1)$-gauge theory of a scalar charged particle on a torus 
$\mathbb{T}^n=\mathbb{R}^n/\Lambda$ to the  analogous theory on 
$\mathbb{R}^n$  constrained by {\it
quasiperiodicity} under translations in the
lattice $\Lambda$. The latter theory provides a {\it
global} description of the former: the quasiperiodic wavefunctions $\psi$ defined
on $\mathbb{R}^n$ play the role of sections of the associated
hermitean line bundle $E$ on $\mathbb{T}^n$, since also $E$ admits a 
global description as a quotient. The components of the covariant
derivatives corresponding to a constant (necessarily integral)
magnetic field $B=dA$ generate a Lie algebra $\gQ$ and together with
the periodic functions the {\it algebra of observables} $\OQ$. The
non-abelian part of $\gQ$ is a Heisenberg Lie algebra with the
electric charge operator $Q$ as the central generator; the
corresponding Lie group $\GQ$ acts on the Hilbert space as the
translation group up to phase factors. Also the space of sections of
$E$ is mapped into itself by $g\in\GQ$. We identify the socalled
magnetic translation group as a subgroup of the {\it observables'
group} $\YQ$. We determine the unitary irreducible representations
of $\OQ,\YQ$ corresponding to integer charges and for each of them
an associated orthonormal basis explicitly in configuration space.
We also clarify how in the $n=2m$ case a holomorphic structure and
Theta functions arise on the associated complex torus.

These results apply equally well to the physics of charged scalar
particles on $\mathbb{R}^n$ and on $\mathbb{T}^n$ in the presence of
periodic magnetic field $B$ and scalar potential. They are also
necessary preliminary steps for the application to these theories of the 
deformation procedure induced by {\it Drinfel'd twists}.

\end{abstract}

\section{Introduction}

Tori are among the simplest manifolds with nontrivial topology.
Gauge theories on them have deserved a lot of attention for both
mathematical and physical reasons.
As known, any $n$-dimensional (real) torus $\mathbb{T}^n$ can be 
described in a global way (i.e. without introducing local trivilizations)
as the quotient $\mathbb{T}^n=\mathbb{R}^n/\Lambda$ of $\mathbb{R}^n$
(the universal cover of the torus) over a lattice 
$\Lambda\subset \mathbb{R}^n$ of rank $n$.
The physical and mathematical communities seem not equally aware
that also gauge theories on tori (i.e. fiber bundles on $\mathbb{T}^n$)
can be described in a global way  as quotients
of gauge theories with trivial topology  (i.e. quotients
of trivial fiber bundles) on $\mathbb{R}^n$.
The first purpose of this paper is to present this fact in elementary terms
sticking for simplicity to the $U(1)$-gauge theory of a scalar quantum particle
(but keeping $n$ and the gauge generic):
we show that on  $\mathbb{T}^n$ such a theory is in fact equivalent to the analogous
(and much simpler) theory on $\mathbb{R}^n$ constrained by
the requirement that under translations by a $\lambda\in\Lambda$
the wavefunctions $\psi(x)$ be quasiperiodic, i.e. periodic up to a $x$-dependent
factor $V$; in more mathematical terms, we show
that the space of sections of a hermitean line bundle $E$ on
$\mathbb{T}^n$ can be described
as a quasiperiodic subspace $\X^V\subset C^\infty({R}^n)$ of the space of
sections of the (topologically trivial) hermitean line bundle on $\mathbb{R}^n$,
modulo gauge transformations on $\mathbb{R}^n$.
The second purpose is to  exploit this equivalence to determine the symmetries 
and give a rather explicit and detailed description of  the irreducible unitary
representations of these theories.

Scalar $U(1)$-gauge theory are physically interesting in themselves,
beside  being propaedeutical to gauge theories
with more complicated gauge groups, particles with spin, etc.
The Schr\"odinger equation for a scalar particle with electric charge $q$
\be
\ba{c} \Ha\psi=i \partial_t
\psi,\qquad \qquad \Ha:=\frac{1}{2m}\nabla\!_a
\nabla\!_a\!+\!{\rm V}, \qquad \qquad
\nabla\!_a:=-i\partial_a\!+\! qA_a,        \ea \label{1Schr}
\ee
(in units such that $\hbar=1$) when considered on $\mathbb{T}^2$ can  describe the dynamics of
such a particle confined on a very thin domain (within the physical space
$\mathbb{R}^3$) modelled as a 2-torus
(a preliminary step to study e.g. the Quantum Hall effect on it).
Formulating (\ref{1Schr}) on $\mathbb{T}^3$ instead of $\mathbb{R}^3$
is a way to obtain a relatively easy finite-volume
theory before e.g. taking the infinite-volume limit.
Similarly, $S=(\psi,\Ha\psi)$ appears as the Euclidean action
in the path-integral quantization of a
charged scalar field in an electromagnetic background
and one may wish to compactify the 4-dimensional Euclidean
spacetime into a torus  $\mathbb{T}^4$ as an infrared cutoff.
In string and Kaluza-Klein theories \cite{GreSchWit88,Pol98},
with or without $D$-branes, 
the same action on tori $\mathbb{T}^n$
of various (usually even) dimensions resulting from the
compactification of the extra-dimensions may appear as part
of the total action.
On the other hand, the equivalent quasiperiodic theories on $\mathbb{R}^n$
are characterized by periodic $B$ and
scalar potential ${\rm V}$, and correspondingly
$\Ha$ becomes the Hamiltonian of Bloch electrons in a magnetic field
$B$. The corresponding dynamical evolution maps $\X^V$ into itself.

The plan of the work is as follows. In section \ref{quasiperiodic}
we introduce {\it quasiperiodicity factors} $V$, spaces
$\X^V$ and show that a compatible connection $A$ on $\mathbb{R}^n$
necessarily yields a periodic curvature 2-form $B=dA$ with
(up to a factor) integer-valued fluxes $\phi_{ab}$
through the basic plaquettes of the lattice (on $\mathbb{T}^n$
these integers will represent the Chern numbers). In the
proof one has to use a self-consistency condition for $V$; on
$\mathbb{T}^n$ the latter becomes the cocycle condition for the
transition functions of a hermitean line bundle  (section \ref{VtoE}).
$V$ is determined by the  $\phi_{ab}$ up to gauge transformations,
and is necessarily a phase factor.  One can decompose
the covariant derivative as
$\nabla=\nabla^{(0)}+QA'$, where $\nabla^{(0)}$
yields a constant $QB\propto [\nabla^{(0)},\nabla^{(0)}]$ and
$A'$ is periodic.
The components $p_a\equiv \nabla^{(0)}_a$
generate a Lie algebra $\gQ$ and together
with the periodic functions the {\it algebra of observables} $\OQ$,
whose elements map $\X^V$ into itself. In particular,
$\Ha,\nabla\!_a\in \OQ$.
The non-abelian part of $\gQ$ is a Heisenberg Lie algebra
with the electric charge operator $Q$ as the central generator.
The corresponding Lie group $\GQ$
acts unitarily on $\X^V$ and its Hilbert space completion $\H^V$
as the translation group
up to phase factors; the latter are trivial if $V=1$, i.e. if the $q\phi_{ab}$
vanish. For these reasons we call $\GQ$ the {\it projective translation
group}. We also introduce
a larger group $\YQ$, which we call {\it observables' group},
of unitary operators on $\H^V$ including 
the functions $e^{il\cdot x}$ ($l\in\mathbb{Z}^n$) as well.
In section \ref{Irreps} we replace $Q\to q\in\mathbb{Z}$ (integer charge)
and identify the   magnetic translation group $M$ of \cite{Zak} as
the discrete subgroup of $\YQ$ commuting with $\GQ$ and
the discrete part of the centre of  $\YQ$  as a subgroup of $M$. 
Then we classify the unitary irreducible
representations of $\OQ,\YQ$ with integer charge and determine them
rather explicitly in configuration space (we write down orthonormal
bases of eigenfunctions of a complete set of commuting observables in
$\OQ$); for
fixed  Chern numbers they are parametrized by a point $\tilde\alpha$
in the reciprocal torus of  $\mathbb{T}^n$ (the `Jacobi torus', or
`Brillouin zone').
Each representation of $\YQ$ extends to the Hilbert space completion $\H^V$
of $\X^V$;  $\H^V$ carries also the associated
representation of the group algebra $\YYQ$ of $\YQ$ (a $C^*$-algebra). 
We also clarify how in the $n=2m$ case a holomorphic structure and Theta
functions arise on the associated complex torus.
In  section \ref{VtoE} we show how to realize a hermitean line bundle
$E\stackrel{\pi}{\mapsto}\mathbb{T}^n$ as a quotient
$E=(\mathbb{R}^n\times \b{C})/\mathbb{Z}^n$, where the free abelian action of
$\mathbb{Z}^n$ on $\mathbb{R}^n\!\times\! \b{C}$
involves the quasiperiodicity factor $V$; we then construct a 
one-to-one correspondence between 
$(\X^V,\nabla)$ and pairs
$\big(\Gamma(\mathbb{T}^n,E),\bnabla\big)$, where $\Gamma(\mathbb{T}^n,E)$
stands for the space of sections of $E$ and  $\bnabla$
the associated covariant derivative.
In general {\it globally defined} gauge transformations on $\mathbb{R}^n$
induce {\it locally defined} gauge transformations on $\mathbb{T}^n$.
$\Gamma(\mathbb{T}^n,E)$ becomes a $\GQ$-module upon lifting
of the action from $\X^V$.
In section \ref{conclu} we summarize the main conclusions of the work.

\medskip
The subject is not new. That unitary representations of the covering group of 
the universal cover of a manifold transform trivial bundles on the former and their sections
into bundles on the latter and their sections is  explained e.g. in  \cite{Sun89,Gru00}.
This is applied to the case of constant magnetic field on  $\mathbb{T}^2,\b{R}^2$ and
 $\mathbb{T}^n,\b{R}^n$ resp. in \cite{BruSun94,Tan}.  
In \cite{AscOveSei94,Gru00} it is shown (for $n=2$ and general $n$,
respectively) that the Hilbert space of states of the theory on  $\b{R}^n$
with a given periodic magnetic field $B$
is the direct integral of the Hilbert spaces of all the inequivalent theories on 
$\mathbb{T}^n$ characterized by the same $B$. 
In \cite{Got95} it is shown that $U(1)$ gauge covariant quantum mechanics 
on a torus is  - in the sense of geometric quantization \cite{Kir90,Woo92} - an example of a full 
quantization of a symplectic manifold, and not only a prequantization.
Several results in sections \ref{quasiperiodic}, 
\ref{VtoE} are the analog in the smooth framework of known results in
the holomorphic one (see e.g. \cite{BirLan04,Mum83}),
but our treatment is based only on the basic definition of a real
torus; additional structures such as
holomorphic ones on complex tori (which
can arise only for $n=2m$), complex
abelian varieties, etc., are not necessary.
The clear definition and decomposition of the algebra/group
of observables $\OQ,\YQ$, the physical identification
of the central generator $Q$ as the electric charge operator,
the complete classification of the irreducible unitary
representations of $\YQ$, as well as an explicit determination
of the latter in configuration space, up to our knowledge are new.
Moreover, our treatment is valid not only
for all $n$ but also for {\it all} gauges.
This is not the case in \cite{Tan} and is not evident e.g. in \cite{BirLan04,Mum83},
where the presence of the gauge group is hidden by the fact that automorphy factors,
that play a role similar to our quasiperiodicity factors $V$,
may take values in all of $\b{C}\setminus\{0\}$ rather than
in the gauge group $U(1)$].

The above results, beside their own interest, provide a setting which allows to
apply the socalled twist-induced deformation procedure to the quantum physics of charged scalar
particles on $\mathbb{R}^n$ and on $\mathbb{T}^n$ in the presence of
periodic magnetic field $B$ and scalar potential: the {\it Drinfel'd twist} $\F$\cite{Dri83},
which is the fundamental object on which the deformation is based, has to be a 
formal power series in the deformation parameter $\lambda$ with
coefficients in $U\gQ\ot U\gQ$, which {\it globally} act on the bundles,
rather than in the abelian algebra $U\g_0\ot U\g_0$, which may only {\it locally} act on the bundles.
For a single particle the deformation has been realized in \cite{Fio11} with Drinfel'd twists of abelian type; with Drinfel'd  twists $\F\in (U\gQ\ot U\gQ)[[\lambda]]$ of general type
(possibly leading even to strictly quasitriangular Hopf algebras) it is current object of work.
Application to many-particle systems and quantum field theories can be then developed
along the lines of \cite{Fio10}.

\medskip
For simplicity we shall assume $\Lambda=2\pi\mathbb{Z}^n$,
whence the reciprocal lattice is $\mathbb{Z}^n$. This is no loss of
generality as far as we are concerned,
since $\Lambda$ can be always transformed into $2\pi\mathbb{Z}^n$
by a linear transformation of $\mathbb{R}^n$,
$x\mapsto gx$, $g\in GL(n)$\footnote{As the holomorphic structure
w.r.t. the complex variables $z^j=x^j\!+\!ix^{m\!+\!j}$ is {\it not}
invariant under  $x\mapsto gx$ for generic $g\in GL(n)$,
the choice  $\Lambda=2\pi\mathbb{Z}^n$ would be a loss of generality
if $n=2m$ and we were concerned with holomorphic line bundles on the complex 
$m$-torus  $\mathbb{T}^n=\mathbb{C}^m/\Lambda$. 
See also the end of section \ref{Irreps}.}. We shall use the following
abbreviations. $\mathbb{N}_0=\mathbb{N}\cup\{0\}$;
$K^t$ stands for the transpose of a matrix $K$;
elements $h,k\in\b{C}^n$ are considered as columns;
$h\cdot k:=h^tk$ (at the rhs the product is row by column);
$u^l\!:=\!e^{il\cdot x}$ for all $l:=(l_1,...,l_m)\in \mathbb{Z}^n$;
$U(1)$ stands for the group of complex numbers of modulus 1;
we denote as $\mathcal{Z}(\A)$ the center of a group
or an algebra $\A$. We denote as $\X$ the subalgebra of
$C^\infty(\mathbb{R}^n)$ consisting of periodic functions $f$,
namely such that  $f(x\!+\!2\pi l)=f(x)$ for all $l\in \mathbb{Z}^n$,
or equivalently of functions (Laurent series) of
$u\equiv(u^1,...,u^n)\equiv(e^{i x^1},...,e^{i x^n})$ only.
$\X$ can be identified also with $C^\infty(\mathbb{T}^n)$.
We denote as $\{e_1,...,e_n\}$
the canonical basis of $\mathbb{R}^n$: $e_1\!:=\!(1,0,...,0)$, etc.


\section{Quasiperiodic wavefunctions and related
connections on $\mathbb{R}^n$}
\label{quasiperiodic}

The simplest way to impose invariance of the particle probability density
under discrete translations $\lambda\in\Lambda$ is to impose it on
the wavefunction $\psi$, i.e. to require it to be periodic.
But this is not necessary; it suffices to require $|\psi|$
to be periodic, that is $\psi$ to be quasiperiodic, i.e.
invariant up to a phase factor $V$.
A set of quasiperiodicity conditions of the form
\be
\psi(x\!+\!2\pi l)=
V\!(l,x)\,\psi(x)\qquad \qquad \forall\, x\in \mathbb{R}^n,
\quad l\in \mathbb{Z}^n              \label{quasiperiodicity}
\ee
relates the values of $\psi$ in any two points $x,x\!+\!2\pi l$
of the lattice $x\!+\!2\pi\mathbb{Z}^n$ through a phase factor
$V\!(l,x)$\footnote{Actually it is not necessary to
assume from the start that $V\in U(1)$; assuming just that it is
nonvanishing complex, $V\in U(1)$ will follow from the reality
of $A_a$, see below.}.
For a Bloch electron in a crystal in absence
of magnetic field a factor depending only on $l$ is
enough: $V=e^{i k\cdot l}$, where $k$ is the {\it quasimomentum}
of the electron. But one can allow  $V$ to depend also on $x$.
In any case nontrivial solutions $\psi$ of (\ref{quasiperiodicity})
may exist only if  the factors relating three generic
points $x,x\!+\!2\pi l,x\!+\!2\pi (\!l \!+\!l'\!)$ of the lattice
are consistent with each other, i.e.
\be
V(l \!+\!l',x)=V(l,x\!+\!2\pi l') \, V(l',x),\qquad \qquad\forall\,
\: l,l'\in \mathbb{Z}^n.                      \label{U1cocycle}
\ee
Note that this implies $V(0,x)\equiv 1$ and
$[V(l,x)]^{-1}=V(-l,x\!+\!2\pi l)$.
We introduce an auxiliary Hilbert space $\HQ$ with
an orthonormal basis $\{|q\ra\}_{q\in\mathbb{Z}}$ and on $\HQ$
a self-adjoint operator $Q$ defined by $Q|q\ra=q|q\ra$.
We regard a smooth wavefunction $\psi$ of a particle
with electric charge $q$ (in $e$ units) as an element
of $C^\infty(\mathbb{R}^n)\ot |q\ra$. As the latter is an eigenspace
with eigenvalue $q$ of $\1\ot Q$, we shall adopt  $\1\ot Q$
as the electric charge operator.
We give the covariant derivative a form independent of $q$
through $\nabla:=(-i)d\ot \1\!+\!A(x)\ot Q$; here $d$ stands
for the exterior derivative. When we risk no
confusion we shall abbreviate $\nabla=-id\!+\!A(x) Q$, \
$\psi\!\in\! C^\infty(\mathbb{R}^n)|q\ra$, etc.
Given a smooth function $V:\mathbb{Z}^n\times \mathbb{R}^n\mapsto
\b{C}\setminus\{0\}$ fulfilling
(\ref{U1cocycle}) consider the space
\be
\X^V:=\{\psi\in C^\infty(\mathbb{R}^n)\ot |q\ra\quad|\quad
\psi(x\!+\!2\pi l)=V\!(l,x)\,\psi(x)\qquad \forall\, x\in \mathbb{R}^n\!, \:
l\in \mathbb{Z}^n\}.
\ee
We look for covariant derivatives $\nabla$
such that their components map $\X^V$ into itself,
\be
\nabla\!_a:\X^V\mapsto\X^V.    \label{connection}
\ee
Given such a $\nabla$, also \
$QB_{ab}(x)\psi(x)=\{\frac i2[\nabla\!_a,\nabla\!_b]\psi\}(x)$ \
fulfills (\ref{quasiperiodicity}), implying
that all the $B_{ab}=\frac 12(\partial_a A_b-\partial_bA _a)$
are periodic functions. From the Fourier expansions
it follows
\be
B_{ab}(x)=\bA _{ab}+
\underbrace{\sum_{l \neq 0}\beta^l_{ab}e^{il\cdot x}}_{B_{ab}'(x)}
\qquad\quad \Rightarrow\quad \qquad A_a(x)=  x^b\bA _{ba}
+\alpha_a+\underbrace{\sum_{l \neq 0}\alpha^l_ae^{il\cdot x}}_{A'_a(x)}
\label{Adeco}
\ee
up to a gauge transformation,
where the periodic 
1-form
 $A'(x)$ is such that $B'=dA'$.
A dependence of $\psi,A',V$ also on the time variable $t$
is not excluded, but we will not write the argument $t$ explicitly.
The reasons why we have isolated the constant parts
$\bA _{ab},\alpha_a\in \mathbb{R}$ in the expansions of
$B_{ab},A_a$ will become clear below.
We decompose the covariant derivative in a gauge-independent
part $A_a'Q$ and a gauge-dependent part $p_a$:
\be
\nabla\!_a:=\!-i\partial_a\!+\!A_aQ=p_a\!+\!A_a'Q,\qquad\qquad
\qquad  A_a'\!\in\!\X;                   \label{decoD}
\ee
$p_a\!=\!-i\partial_a\!+\!x^b\bA _{ba}Q\!+\!\alpha_a Q$
up to a gauge transformation.
Going back to (\ref{connection}), $\nabla\!_a\psi$ will fulfill
(\ref{quasiperiodicity}) iff also $p_a\psi$ does,
by the periodicity of $A'_a(x)$;
up to a gauge transformation this yields\footnote{ ~{}
\vskip-1truecm
\bea
&&0\stackrel{!}{=}(p_a\psi)(x\!+\!2\pi l)-V\!(l,x)\,(p_a\psi)(x)
=(-i\partial_a\!+\!x^b\bA _{ba}q\!+\!2\pi l_b\bA_{ba}q\!+\!
\alpha_aq)V\!(l,x)\,\psi(x)\nn[6pt]
&&\qquad-(-i\partial_a\!+\!x^b\bA _{ba}q\!+\!\alpha_aq) V\!(l,x)\,(\psi)(x)
=[(-i\partial_a\!+\!2\pi l_b\bA_{ba}q)V\!(l,x)]\,\psi(x)
\qquad\quad\Rightarrow\quad\qquad (\ref{gauge0})_1. \nonumber
\eea
}
the first formula in
\be
V(l,x)\equiv V^{\bA }(l,x)=e^{-iq2\pi l^t\!\bA \! x},\qquad\qquad
p_a\!=\!-i\partial_a\!+\!x^b\bA _{ba}Q\!+\!\alpha_a Q,
                                  \label{gauge0}
\ee
which is consistent with (\ref{U1cocycle}) for all
eigenvalues $q\in\mathbb{Z}$ of $Q$ iff
the quantization conditions
\be
\nu_{ab} \in\mathbb{Z},\qquad\qquad
\nu_{ab}:=2\pi\bA _{ab}
                              \label{quantizedB}
\ee
for all $a,b$ are satisfied\footnote{As \
$V^{-1}(l \!+\!l',x)V(l,x\!+\!2\pi l') V(l',x)=
e^{-iq4\pi^2l^t\!\bA \!l'}$, \ (\ref{U1cocycle}) amounts to \
$q2\pi l^t\bA l'\in \mathbb{Z}$ \ for all \ $q\in \mathbb{Z}$ and $l,l'\in \mathbb{Z}^n$,
\ i.e. to (\ref{quantizedB}).}. (This in particular excludes a dependence
of $\bA$ on $t$).
For $q=0$ or $\bA=0$ we find $V\equiv 1$ and $\X^1=\X\ot|0\ra\simeq\X$.
Otherwise (\ref{quasiperiodicity}), (\ref{gauge0})$_1$
do not admit solutions of the form $\psi(x)=f(x)$, $f\in\X$.
Introducing fundamental $k$-dimensional  cells $C_{a_1...a_k}^y$
for $k\le n$ and $a_1<a_2<....< a_k$ by
\be
C_{a_1...a_k}^y:=\{x\in\mathbb{R}^n \:\: |\:\: x^{a_h}\in
[y^{a_h},y^{a_h}\!+\!2\pi[, \: h=1,...,k; \:\: x^a=y^a\:
\mbox{otherwise}\},                        \label{fundamentalcells}
\ee
one easily finds that the flux
$\phi_{ab}$ of $B=B_{ab}dx^adx^b$  through a plaquette $C^y_{ab}$
equals that of $\bbA=\bA_{ab}dx^adx^b$
\be
\phi_{ab}=\int_{C_{ab}^y}\!\!\!  B=\int_{C_{ab}^y}\!\!\! \bbA
=2\pi\nu_{ab}             \label{fluxes}
\ee
and more generally\footnote{In fact,
\bea
\int\limits_{C_{a_1...a_{2m}}^y}\!\!\! \!\!\! B^m
\stackrel{(\ref{Adeco})}{=}
\int\limits_{C_{a_1...a_{2m}}^y}\!\!\! \!\!\! [\bbA
\!+\!dA']B^{m\!-\!1}
=\int\limits_{C_{a_1...a_{2m}}^y}\!\!\!\! \!\!\! \bbA B^{m\!-\!1}
+\int\limits_{C_{a_1...a_{2m}}^y}\!\!\!\! \!\!\!d(A'B^{m\!-\!1})
=\int\limits_{C_{a_1...a_{2m}}^y}\!\!\! \!\!\! \bbA B^{m\!-\!1}
+\int\limits_{\partial C_{a_1...a_{2m}}^y}\!\!\!\!\!\!
\!\!\! A'B^{m\!-\!1}\nn
=\int\limits_{C_{a_1...a_{2m}}^y}\!\!\!\!\! \!\!\! \bbA B^{m\!-\!1}=...=
\int\limits_{C_{a_1...a_{2m}}^y}\!\!\! \!\!\! (\bbA)^m=
\int\limits_{C_{a_1...a_{2m}}^y}\!\!\! \!\!\!\bA_{[a_1a_2}
...\bA _{a_{2m\!-\!1}a_{2m}]}
dx^{a_1}...dx^{a_{2m}}\stackrel{(\ref{quantizedB})}{=}(2\pi)^m
\,\nu_{[a_1a_2}\nu_{a_{2m\!-\!1}a_{2m}]}\nonumber
\eea
The second equality holds because $dB=0$, the third by Stokes theorem,
the fourth by the periodicity of $A',B$, which makes the border integral vanish.
}
\be
\int_{C_{a_1...a_{2m}}^y}\!\!\! \!\!\! B^m=
\int_{C_{a_1...a_{2m}}^y}\!\!\! \!\!\! (\bbA)^m
=(2\pi)^m\nu_{[a_1a_2}\nu_{a_{2m\!-\!1}a_{2m}]}          \label{fluxesm}
\ee
for all $m\le n/2$ and $a_1<a_2<....< a_{2m}$;
the square bracket denotes antisymmetrization w.r.t.
the indices $a_1a_2... a_{2m}$. The results are independent of $y$.
From (\ref{gauge0}-\ref{quantizedB}) we easily find that
\be
\frac i{4\pi}\log\left[\frac {V(l,x\!+\!2\pi l') V(l',x)}
{V(l',x\!+\!2\pi l) V(l,x)}\right]=q l^t\nu l'=:\tilde\nu(l,l')
                                       \label{sigmachern1}
\ee
is $x$-independent and defines an antisymmetric bilinear
$\mathbb{Z}$-valued form $\tilde\nu$ on $\Lambda=\mathbb{Z}^n$; later we shall see
that it is also gauge-independent. Hence
$2\pi\bA=\nu$ can be recovered from the quasiperiodicity factor $V$, in any gauge.

As side-remarks we note that:
1. Relation (\ref{gauge0})$_1$ implies that $V(l,x)$ could not be
an arbitrary complex number: it {\it must} be $V(l,x)\in U(1)$ in the gauge
under consideration and, by (\ref{gaugetr})$_3$, in all gauges.
2. The $V^{\bA}$ make up a (discrete) abelian group
upon pointwise moltiplication $V^{\bA}V^{\bA{}'}=V^{\bA\!+\! \bA{}'}$,
and similarly the $\psi^{\bA}$, but we will not use this fact here.
3. Rational values for the expressions
$2\pi\bA _{ab}=\frac 1{2\pi}\phi_{ab} $
could be obtained imposing quasi-periodicity on some
$k$-fold enlarged lattice ($k\in\mathbb{N}$), as done e.g. in
\cite{Gru00,MorPol01}\footnote{For instance, imposing the conditions
(\ref{quasiperiodicity}-\ref{U1cocycle}) for all $q\in\mathbb{Z}$ and
only for $l,l'$ such that $l_1=kh_1$, $l_1'=kh_1'$ (with
some $h_1,h_1'\in\mathbb{Z}$) would lead to
$k\nu_{1b},  k\nu_{b1} \in\mathbb{Z}$
i.e. $\nu_{1b},\nu_{b1}\in\mathbb{Q}$, and again to $\nu_{ab} \in\mathbb{Z}$
for $a,b>1$.
The corresponding $\psi$'s could be interpreted as  $k$-component
wavefunctions, i.e. sections of a $\b{C}^k$-vector bundle.}.

Summing up, the (left) action of $Q,p_a$ and multiplication
(from either side) by any $f\in\X$ map $\X^V$ into itself.
In other words, $\X^V$ is a $\X$-bimodule and a (left) module of the
$*$-algebra $\OQ$ of smooth differential operators that are  polynomials
in $Q,p_1,...,p_m$ with  coefficients $f$ in $\X$,  constrained by
\be
\ba{lll} [p_a,p_b]=-i2\bA _{ab}Q, \qquad\qquad & [Q,\cdot\:]=0,
\qquad\qquad  &[p_a,f]=-i(\partial_a f),\\[9pt]
f^*(x)=\overline{f(x)}, \qquad\qquad   &p_a^*=p_a,\qquad\qquad  &Q^*=Q.
\ea        \label{comrelO}
\ee
Note that these defining relations of $\OQ$ depend  on the $A_a$ only
through the $\bA _{ab}$ of (\ref{quantizedB}), so are gauge-independent.
In particular $\nabla\!_a,\Ha\in\OQ$.

$Q,p_a$ generate a real  Lie algebra $\gQ$ which is a (non-abelian)
central extension of the abelian Lie algebra $\mathbb{R}^n$ spanned by
the $-i\partial_a$; the
$-2\bA _{ab}$ play the role of structure constants.
$\OQ$ and $\X$ are $U\gQ$-module $*$-algebras under the action
\be
\ba{ll}
 p_a\trc p_b=-i2\bA _{ab}Q, \qquad\qquad & p_a\trc f=-i(\partial_a f),\\[8pt]
Q\trc f=0, \qquad\qquad  & Q\trc p_a=0,  \label{gaction'}
\ea
\ee
for all $f\!\in\!\X$, and
$\X^V$ is a left $U\gQ$-equivariant $\OQ$-module and $\X$-bimodule
(but not an algebra, unless $V\!\equiv\! 1$); this means that
all these structures are compatible with each other and the Leibniz
rule\footnote{Namely, for all $c\in\OQ$, $\psi\in\X^V$, $f\in\X$, $g\in\gQ$
$g\trc (c\psi f)=(g\trc c)\psi f\!+\!c(g\trc \psi) f\!+\!c\psi (g\trc f)$.
}.
`Exponentiating' $\gQ$ we obtain a Lie group
$\GQ$  of transformations
$g_{\tilde z}:\X^V\mapsto\X^V$, \
$\tilde z\equiv(z^0,z)\in\mathbb{R}^{n\!+\!1}$:
\be
g_{\tilde
z}\!=\! e^{i\left[p\cdot z+Q z^0\right]}; \qquad \qquad
g_{\tilde z}g_{\tilde z'}=g_{\tilde z \cp \tilde z'},
\qquad\qquad \tilde z \cp \tilde z':=\tilde z \!+\! \tilde z'\! -\!(z^t\bA
z',0,...,0).        \label{GQlaw}
\ee
This group law formally follows
from the Baker-Campbell-Hausdorff formula\footnote{
\be
e^Re^S=e^{R\!+\!S}
e^{\frac 12[R,S]},
\qquad\qquad \mbox{ if }\:
[R,S]\: \mbox{ commutes with }\:  R,S.  \label{BCH}
\ee }. In the gauge
(\ref{gauge0}) the actions of $Q,p_a$ and $g_{\tilde z}$ read
\be
\ba{l}
Q\!\trc\!\psi\!=\!q\psi,\qquad  p_a\!\trc\!\psi\!=\!(-i\partial\!+\!
qx^t\bA
\!+\!q\alpha\! )_a\psi,\\[8pt]
 [g_{\tilde
z}\!\trc\!\psi](x)=e^{iq\!\left[z^0\!+x^t\!\bA z\!+\alpha^tz\!\right]}
\psi(x\!+\!z);
\ea                         \label{GQaction}
\ee
as a consistency check one can verify the group law (\ref{GQlaw}) and
that $g_{\tilde z}\trc\psi$ indeed fulfills
(\ref{quasiperiodicity}). We shall call $\GQ$ the {\it projective
translation group} because it acts as the abelian group $G_0$ of
translations $\psi(x)\mapsto \psi(x\!+\!z)$ followed by
multiplication by a phase factor, which is necessary to obtain
again a wavefunction in $\X^V$.
The phase factor reduces to 1 only  for $\bA=0$ or on carrier spaces
characterized by $q=0$, i.e. on $C^\infty(\mathbb{R}^n),\X$;

Let $L(n,\mathbb{Z})$ be the group of matrices with integer entries and
determinant equal to $\pm 1$, $r$ the integer
$r:=\frac 12\mbox{rank}(\bA )$
(we recall that the rank of an antisymmetric matrix is always even);
clearly $0\le 2r\le n$.
By the Frobenius theorem (see e.g. \cite{Igu72}, p. 71) there exists
a $S\in L(n,\mathbb{Z})$
such that
\be
\bA =S^t\bar\bA  S,\qquad \qquad\bar\bA :=\left(\ba{ccc}
   & -b  &            \\
b   &    &           \\
    &    &  0_{n\!-\!2r}
\ea \right),\qquad b:=\mbox{diag}(b_1,...,b_r), \label{Frobenius}
\ee
where
$(\nu_1,...,\nu_r):=(2\pi b_1,...,2\pi b_r)$ is a sequence of positive
(or negative) integers
such that each $\nu_{j\!+\!1}$ is an integer multiple of $\nu_j$ [e.g.
$(\nu_1,...,\nu_4)=(3, 6,18,18)$], $0_{n\!-\!2r}$ is the zero
$({n\!-\!2r})\!\times\!({n\!-\!2r})$ matrix
and all the missing blocks are zero matrices of the appropriate sizes.
Therefore, after the change of generators
\be
p_a\mapsto (S^tp)_a,\qquad x^a\mapsto (S^{-1}
x)^a,\qquad \Rightarrow\qquad u^l\mapsto
u^{(S^{-1})^tl},  \label{modular}
\ee
resp. in $\gQ$, $C^\infty(\mathbb{R}^n)$ and $\X$,
the relations (\ref{comrelO}) will
involve the matrix $\bar\bA$ instead of the original $\bA$;
more explicitly,
(\ref{comrelO})$_1$ will become the commutation relations
\be
[p_j, p_{r\!+\!j}]=i2b_jQ \qquad j=1,...,r, \qquad\qquad [p_a,p_b]=0
\qquad \mbox{otherwise}. \label{comrelp}
\ee
This shows that
\be
\gQ\simeq \heQ{}_{2r\!+\!1}\oplus\mathbb{R}^{n\!-\!2r},\qquad\qquad
\GQ\simeq \HeQ{}_{2r\!+\!1}\times\mathbb{R}^{n\!-\!2r}; \label{decomp}
\ee
$\heQ{}_m,\!\HeQ{}_m$  denote the
Heisenberg Lie algebra, group of dimension $m$ and central generator $Q$.

\medskip
The Weyl forms of $x^a{}^*=x^a$, (\ref{comrelO})
and its consequence $[p_a,x^b]=-i\delta^b_a$,
are easily
determined with the help of (\ref{BCH}) and synthetically read
\be
\ba{l}
e^{i(h\cdot x+p\cdot y+Qy^0)} e^{i(k\cdot x+p\cdot z+Qz^0)}
=e^{i\left[(h+k)\cdot x+p\cdot (y+z)+Q(y^0+z^0)\right]}\,
e^{\frac i2\left[k\cdot y-h\cdot z+2Q y^t\bA  z\right]}
\\[9pt]
\left[e^{i(h\cdot x+p\cdot y+Qy^0)} \right]^*=
e^{-i(h\cdot x+p\cdot y+Qy^0)}
\ea            \label{weyl}
\ee
for any $h,k\in\mathbb{R}^n$ and $(y^0,y),(z^0, z)\in\mathbb{R}^{n\!+\!1}$.
We define groups $\YQ,\GQ,L$ by
\be
\ba{l}
\YQ:=\{  e^{i(l\cdot x+p\cdot z+h^0+Qz^0)}\:\: |
\:\:h^0\in\mathbb{R},\:\: l\in\mathbb{Z}^n,\:\: (z^0,z)\in\mathbb{R}^{n\!+\!1}\},\\[8pt]
\GQ:=\{ e^{i(p\cdot z+Qz^0)}\:\: |\:\:
(z^0,z)\in\mathbb{R}^{n\!+\!1}\},\\[8pt]
 L:=\{e^{i(l\cdot x+h^0)}\:\: |
\:\: l\in\mathbb{Z}^n, \:\:h^0\in\mathbb{R}\};\ea                \label{defvarie}
\ee
the group laws can be read off (\ref{weyl})
and depend only on $\bA$.
$L$ is isomorphic to $\mathbb{Z}^n\times U(1)$ and is a normal subgroup
of $\YQ$; actually  $\YQ$
is the semidirect product $\YQ= \GQ\cross  L$.
All elements of $\YQ$ are unitary
w.r.t. the involution (\ref{weyl})$_2$.
We shall call  $\YQ$ the  {\it observables' group} because
its elements are observables, i.e. map $\X^V$ into itself;
the $\GQ$ part acts as in (\ref{GQaction}), the $L$ part acts by
multiplication. Moreover,
we shall call $\YYQ$ the group algebra of $\YQ$; it is a
$C^*$-algebra.


By (\ref{quasiperiodicity}) $\psi'{}^*\psi$ is periodic
for all $\psi',\psi\in\X^V$, and the formula
\be
(\psi',\psi):=\int_{C^y_{1...n}}\!\!\!\!\!\!d^nx\:\,
\overline{\psi'(x)}\,\psi(x), \qquad
\label{hermstr}
\ee
where $C^y_{1...n}$ is any fundamental $n$-dimensional cell
(\ref{fundamentalcells}) [e.g.
$C^0_{1...n}=\left([0,2\pi[\right)^n$],
defines a hermitean structure in $\X^V$ making the latter a
pre-Hilbert space.
As $p_a\trc (\psi'{}^*\psi)\equiv p_a(\psi'{}^*\psi)
=-i\partial_a(\psi'{}^*\psi)$,
which has a vanishing integral,
by the Leibniz rule the $p_a$ are essentially self-adjoint.
We shall call $\H^V$ the Hilbert space completion of $\X^V$.
$\YQ$ extends as a group of unitary transformations of  $\H^V$;
the $L$ part still acts by multiplication, and the $\GQ$ part
in the above gauge acts still as in (\ref{GQaction})$_3$.

Fixed any function $\psi_0\in\X^V$  vanishing nowhere,
$\psi\psi_0^{-1}$ is well-defined and periodic, i.e. in $\X$,
for all $\psi\in\X^V$,
whence the decomposition $\X^V= \X\, \psi_0$.

Given a $*$-representation $\big(\rho(\OQ),\X^V\big)$ of $\OQ$ as a
$*$-algebra of operators  on the pre-Hilbert space $\X^V$
(by definition $o\trc\psi=: \rho(o)\psi$ for all $o\in\OQ$ and
$\psi\in\X^V$), a unitary equivalent one is obtained
through a smooth gauge transformation $U=e^{iq\varphi}$
[$\varphi(x)$ smooth and real valued] acting as the unitary transformation
$\big(\rho(\OQ),\X^V\big)\mapsto \big(\rho^U(\OQ),\X^{V^U}\big)$
defined by
\be
\rho^U(o)=U \rho(o)U^{-1},\qquad\quad
\psi^U=U\psi,\qquad\quad
V^U(l,x)\!=\!U(x\!+\!2\pi l)\, V(l,x)\, U^{-1}(x).     \label{gaugetr}
\ee
With  \ $o\!\in\!\YQ$ and $\psi\!\in\!,\H^V$ \ (\ref{gaugetr}) defines also
a unitary transformation
$\big(\rho(\YQ),\H^V\big)\mapsto \big(\rho^U(\YQ),\H^{V^U}\big)$.
Although the realization of the $p_a$ is gauge-dependent,
all the relations (\ref{quasiperiodicity}-\ref{connection}),
(\ref{Adeco})$_1$
(\ref{decoD}), (\ref{quantizedB}-\ref{GQlaw}),
(\ref{Frobenius}-\ref{hermstr}) remain valid;
in particular, the $\mathbb{Z}$-valued 2-form $\tilde\nu$
(\ref{sigmachern1}) is gauge-invariant and allows to recover
form the quasiperiodicty factors the integers $2\pi\bA=\nu$.
Calling $\big(\rho(\OQ),\X^V\big)$ the representation
[based on (\ref{gauge0})] that we have used so far, choosing
$U(x)=e^{i \frac q2  x^t\bS x}$ and setting
$\beta:=\bA \!+\!\bS$, we find an equivalent representation
$\big(\rho^U(\OQ),\X^{V^U}\big)$ characterized by
\be
\ba{l}
 V^{U}(l,x)
=e^{-iq2\pi l^t\beta (x\!+\! l\pi)},
\qquad\qquad
p_a=-i\partial_a\!+\!x^b\beta_{ba}Q\!+\!\alpha_a Q
\ea                                \label{quasiper}
\ee
[for $U(x)\equiv 1$, i.e.
$\beta=\bA$, we recover the original gauge (\ref{gauge0})].
We shall adopt the shorter notations $\X^\beta\equiv\X^{V^U}$,
$\H^\beta\equiv\H^{V^U}$, etc. for the spaces of complex functions
fulfilling (\ref{quasiperiodicity}) with $V$ given by (\ref{quasiper}).
Performing a change (\ref{modular}) and choosing $\bS$ so that
$\beta$ becomes lower-triangular we find
\be
\beta\equiv\bar\beta=\left(\ba{ccc}
   &  0_r  &            \\
2b   &    &           \\
    &    &  0_{n\!-\!2r}
\ea \right),
                                        \label{lowertriang}
\ee
and eq. (\ref{quasiperiodicity}) becomes
\be
\psi(x\!+\!2\pi l)=
e^{-i2q\sum\limits_{j=1}^r \nu_j l_{r\!+\!j}x^j}
\,\psi(x)\qquad \qquad \forall\, x\in \mathbb{R}^n,
\: l\in \mathbb{Z}^n.              \label{quasiperiodicityn}
\ee
Choosing $l_{r\!+\!1}=...=l_{2r}=0$ 
one finds that $\psi(x)$ is periodic in
$x^1,...,x^r,x^{2r\!+\!1},...,x^n$. It is straightforward to check
that on the Fourier decomposition of $\psi$ w.r.t. $x^1,...,x^r$
$$
\psi(x)=
\sum\limits_{k\in \mathbb{Z}^{r}}
e^{i\sum\limits_{j=1}^rk_jx^j}\psi_{k}(x^{r\!+\!1},...,x^{n}),
$$
condition (\ref{quasiperiodicityn}) for arbitrary
$l$ reduces to the recurrence relations
$$
\psi_{k\!+\!2q(\nu_1l_{r\!+\!1},...,\nu_rl_{2r})}(x^{r\!+\!1},...,x^n)
=\psi_{k}(x^{r\!+\!1}\!+\!2\pi l_{r\!+\!1},...,
x^{2r}\!+\!2\pi l_{2r},x^{2r\!+\!1},...,x^{n}).
$$
By the latter one can express all $\psi_k$ in terms
of those  with $k\in K$, where
\be
 K:=\{0,1,...,|2q\nu_1|\!-\!1\}\times...\times
\{0,1,...,|2q\nu_r|\!-\!1\}\subset \mathbb{Z}^{r}.
\ee
Therefore the most general solution of (\ref{quasiperiodicityn})  reads
\be
\psi(x)=\sum\limits_{k\in K}\sum\limits_{l\in \mathbb{Z}^r}
e^{i\sum\limits_{j=1}^r(k_j\!+\!2q \nu_j l_j)x^j
}\psi_k(x^{r\!+\!1}
\!\!+\!2\pi l_1,...,x^{2r}\!\!+\!2\pi l_r,x^{2r\!+\!1},...,x^n ).
\label{solutionsn}
\ee
Replacing in (\ref{hermstr}) we find
\be
(\psi',\psi)\, =\, (2\pi)^{r} \,  \sum\limits_{k\in K}
\int_{\mathbb{R}}\!\!dx^{r\!+\!1}...\int_{\mathbb{R}}\!\!dx^{2r}
\int_0^{2\pi}\!\!\!\!\!dx^{2r\!+\!1}...
\int_0^{2\pi}\!\!\!\!\!dx^n\: \psi_k'{}^*\psi_k .
\label{hermstrn}
\ee
Hence \ $\psi\in\H^\beta$
iff \ $\psi_k\in\L^2\!\left(\mathbb{R}^r\!\times\!\mathbb{T}^{n\!-\!2r}\right)$,
\ $\psi\in\X^\beta$ \ iff \ $\psi_k\in
\mathcal{S}\!\left(\mathbb{R}^r\!\times\!\mathbb{T}^{n\!-\!2r}\right)$,
\ for all $k\in K$, and
\be
\H^\beta=\bigoplus\limits_{k\in K} \H^k,\qquad\qquad\qquad
\X^\beta=\bigoplus\limits_{k\in K} \X^k.   \label{direcstsumn}
\ee
where we have denoted as $\X^k\subset\X^\beta$, $\H^k\subset\H^\beta$
the subspaces characterized by $\psi_{k'}\equiv 0$ for
$k'\in K\setminus\{k\}$
and $\psi_k$ belonging to
\ $\L^2\!\left(\mathbb{R}^r\!\times\!\mathbb{T}^{n\!-\!2r}\right)$, \
$\mathcal{S}\!\left(\mathbb{R}^r\!\times\!\mathbb{T}^{n\!-\!2r}\right)$
respectively. The correspondence
$\psi\leftrightarrow \{\psi_k\}_{k\in K}$ is the generalized
Weil-Brezin-Zak transform,
see 
\cite{Zak68} and 1.10 in \cite{Fol89}.
One can easily check that each $\H^k$ is mapped into itself by $G$
(resp. each $\X^k$ is mapped into itself by $U\g$). We shall show
this fact in next subsection while presenting bases of $\X^k,\X^\beta$.

\section{Decomposition and irreducible representations of the
observables' group for integer charge}
\label{Irreps}

Let us denote as $Y,
\Y, G,g,\Oo,{\bf h}_m,H_m$ the
groups/algebras obtained from $\YQ,
\YYQ,\GQ,\gQ,\OQ$, $\heQ{}_m,\HeQ{}_m$
replacing the central element $Q$ in
(\ref{comrelp}-\ref{defvarie}) by 
some
$q\in \mathbb{Z}$.
In this subsection we decompose them and identify
their unitary irreducible representations.
We abbreviate $\tilde A\!:=\!qA$, and similarly for
all the derived objects: $\tilde \alpha\!:=\!q\alpha$, etc.
Note that $q,\bA$ enter the commutation relations
only through their product $\tilde\bA\!:=\!q\bA$.


In (\ref{defvarie}) we can set $z^0=0$, as we can reabsorb $qz^0$
into a redefinition of $h^0$. Define
\be
\mathcal{Z}^{(n)}:=\big\{\zeta^l:=e^{i2\pi l^t(p+ 2\tilde\bA x)} \:\: |\:\: l\in\mathbb{Z}^n
\big\},\qquad\qquad
\zeta_a:= e^{i2\pi(p_a+2\tilde \bA_{ab}x^b)}.
\ee
Using (\ref{weyl})
it is easy to check that $[\zeta^l,e^{i(l'\cdot x+p\cdot z+h^0)}]=0$
for all $l,l'\in\mathbb{Z}^n$,  $(h^0,z)\in\mathbb{R}^{n\!+\!1}$.
Hence $\mathcal{Z}^{(n)}$ is a discrete subgroup of $\mathcal{Z}(Y)$,
isomorphic to $\mathbb{Z}^n$; it is
generated by the $\zeta_a$.

Let $r:=\frac 12\mbox{rank}(\tilde \bA )$. We first
perform a change (\ref{modular}) (chosen so that
$\tilde \nu_j\!=\!2\pi \tilde b_j\!\in\!\mathbb{N}$, i.e. are positive)
and then define
\be
\ba{l} \check p_j\!:=\!2\tilde b_j
x^j\!+\!p_{r\!+\!j},\qquad \check x^j\!:=\!\frac
1{2\tilde b_j}\!p_j\!-\!x^{r\!+\!j},\qquad \check x^{r\!+\!j}\!:=\!\frac
1{2\tilde b_j}p_j,\qquad
\check p_{r\!+\!j}\!:=\!p_{r\!+\!j},\qquad j\!=\!1,\!...\!,\!r\\[8pt]
\check x^a\!:=\!x^a,\qquad \check p_a\!:=\!p_a \qquad\qquad\qquad\qquad a>2r
\ea
\ee
one obtains objects fulfilling the canonical commutation relations \
$[\check x^a,\check p_b]\!=\!i\delta^a_b$, \ 
$[\check x^a,\check x^b]\!=\!0\!=\![\check p_a,\check p_b]$ \ for $a,b\!=\!1,...n$.  The set \
$\{1, \check x^{r\!+\!1},...,\check x^{2r},
\check p_{r\!+\!1},...,\check p_n\}$ is a new basis of $\g$.
The elements
\be
m_j:=e^{i\check x^j}=[u^{r\!+\!j}]^{-1}
e^{\frac {i\pi}{\tilde \nu_j}p_j},\qquad \qquad
m_{r\!+\!j}:=e^{\frac i{2\tilde b_j}\check p_j}=u^j
e^{\frac {i\pi}{\tilde \nu_j}p_{r\!+\!j}}
\ee
commute with $G,\g$ (and hence also with $U\g$) and
fulfill 
\be
m_jm_{r\!+\!j}=m_{r\!+\!j}m_je^{\frac {i\pi}{\tilde \nu_j}},
\qquad j=1,...,r,\qquad\qquad m_am_b=m_bm_a\quad\mbox{ otherwise.}
 \label{comrelm}
\ee
From
$$
l\!\cdot\! x\!+\!p\!\cdot\! z=\sum_{j=1}^r\left[
\frac {l_j}{2\tilde b_j}\check p_j\!-\!l_{r\!+\!j}\check x^j+
p_j\left(z^j\!+\!\frac{l_{r\!+\!j}}{2\tilde b_j}\right)
\!+\! p_{r\!+\!j}\!\left(z^{r\!+\!j}\!-\!\frac
{l_j}{2\tilde b_j}\right)\right]+\sum_{a=2r\!+\!1}^n\!(l_ax^a+p_a z^a)
$$
and (\ref{BCH}) we obtain the following decomposition of the generic element
of $Y$:
\be
\ba{l}
e^{i(l\cdot x+p\cdot z+h^0)}=\left[\prod\limits_{j=1}^r
m_j^{-l_{r\!+\!j}}m_{r\!+\!j}^{l_j}
e^{\frac {i\pi}{2\tilde \nu_j}l_jl_{r\!+\!j}}\right] \,\left[e^{ih^0}\prod\limits_{j=1}^r  e^{ip_j\left(z^j\!+\!\frac {l_{r\!+\!j}}{2\tilde b_j}\right)
\!+\!i p_{r\!+\!j}\!\left(z^{r\!+\!j}\!-\!\frac {l_j}{2\tilde b_j}\right)}\right]
\,\left[\prod\limits_{a=2r\!+\!1}^ne^{i(l_ax^a+p_az^a)}\right].
\label{decoy}
\ea
\ee
For any  hermitean  ${\rm q},{\rm p}$ fulfilling $[{\rm q},{\rm p}]\!=\!i$
denote as  \ 
$H_3({\rm q},{\rm p})\!:=\!\{e^{i(h^0+y{\rm q}+ z{\rm p})}\:\: | \:\: 
(h^0\!,\!y,z)\!\in\!\mathbb{R}^3\}$ \ the associated 3-dimensional Heisenberg Lie group
and  \ ${\rm Y}({\rm q},{\rm p})\!:=\!\{e^{i(l{\rm q}+ z{\rm p}+h^0)}\:\: | \:\: 
(l,h^0\!,\!z)\!\in\!\mathbb{Z}\!\times\!\mathbb{R}^2\}$ (this group plays a role in 
quantum mechanics on the circle, see below).

\begin{prop}
\ \ $Y$ decomposes into a product of  commuting subgroups as follows:
\be
Y= M^1...M^r\: H_{3}^1...H_{3}^r\: Y^{2r\!+\!1}...Y^n.
\label{decoY}
\ee
The  elements of $M^1,...,M^r$, $H_{3}^1,...,H_{3}^r$, $Y^{2r\!+\!1},...,Y^n$
are those ranging resp. in the square brackets of the
1$^{st}$, 2$^{nd}$, 3$^{rd}$ block at the
rhs(\ref{decoy}), for all the possible values of $l,z,h^0$.
Namely,
\begin{itemize}
\item  $M^j$ 
is discrete, generated by
$m_j,m_{r\!+\!j},e^{\frac {i\pi}{\tilde \nu_j}} $ fulfilling
(\ref{comrelm}) and by the inverses $m_j^{-1},m_{r\!+\!j}^{-1}$;

\item $H_{3}^j\!:=\!\{
e^{i(h^0+wp_j+ zp_{r\!+\!j})}\:\: | \:\:  (h^0,w,z)\!\in\!\mathbb{R}^3\}=
H_{3}\left(\frac{p_j}{2\tilde b_j}, p_{r\!+\!j}\right)$;

\item $Y^a:={\rm Y}(x^a,p^a)$.
\end{itemize}

\noindent
Moreover, \ $\mathcal{Z}(Y)=\mathcal{Z}^{(n)}\, U(1)$; \ if we adopt
the $x,p$  variables at the rhs(\ref{modular}) then
$\zeta_j=(m_j)^{2\tilde \nu_j}$, \ $\zeta_{r\!+\!j}=(m_{r\!+\!j})^{2\tilde \nu_j}$
($j=1,...,r$).
\end{prop}

\bp~
The rest of the proof is as follows. The
elements of the different subgroups at the rhs's of the relations
(\ref{decoY}) commute with each other by the relations
$[\check x^a,\check p_b]=i\delta^a_b$. \ Clearly
$\mathcal{Z}(H_{3}^1...H_{3}^r)=\mathcal{Z}(Y^{2r\!+\!1}...Y^n)=
\{e^{ih^0}  \: | \:h^0\!\in\!\mathbb{R}\}=U(1)$,
whence $\mathcal{Z}(Y)=\mathcal{Z}^{(n)\, }U(1)$,
as claimed.
\ep

\noindent
The center $\mathcal{Z}(\Y)$ of the $C^*$-algebra $\Y$
is generated by the $\zeta^a$.
The irreducible unitary representations
(briefly {\it irreps}) of $\Y$ are those of $Y$.
We now study them, starting 
from 
the lowest $n$'s.

\subsubsection*{Irreducible unitary representations of $Y$ for $n=1$}
$\bA\!=\!0$, $r\!=\!0$,
$Y\!\equiv\! Y^1\!\simeq\!{\rm Y}$ and $\Oo$ is the algebra of observables
of quantum mechanics on a circle.
The formulae
\be
\rho_{\tilde\alpha}\big[e^{i(lx+h^0)}\big]\psi(x)= e^{i(lx+h^0)}\psi(x),
\qquad \quad
\rho_{\tilde\alpha}(e^{izp})\psi(x)=
e^{i\tilde\alpha z +z\partial_x }\psi(x)=e^{i\tilde\alpha z}\psi(x\!+\!z),
                                                     \label{irrepsn=1}
\ee
where $\tilde\alpha\in\mathbb{R}$ and $\psi\in\L^2(S^1)$,
define an irrep \ $(\rho_{\tilde\alpha},\H_{\tilde\alpha})$ \ of ${\rm Y}$,
with $\H_{\tilde\alpha}\!=\!\L^2(S^1)$.
The formulae
\be
\rho_{\tilde\alpha}\big[e^{ilx}\big]\psi(x)= e^{ilx}\psi(x),
\qquad \qquad
\rho_{\tilde\alpha}(p)\psi(x)=
(\tilde\alpha  -i\partial_x )\psi(x)               \label{irrepsn=1'}
\ee
define the associated irrep 
$(\rho_{\tilde\alpha},\X_{\tilde\alpha})$ of $\Oo$
on the dense subspace $\X_{\tilde\alpha}=\C^\infty(S^1)\subset\L^2(S^1)$.
If $\tilde\alpha'= \tilde\alpha\!+\!l$ then
$\rho_{\tilde\alpha}$ and $\rho_{\tilde\alpha'}$
are related by the unitary transformation
$\psi(x)\mapsto e^{ilx}\psi(x)$ and therefore are equivalent.
This is in agreement with the fact that the casimir $\zeta=e^{i2\pi p}$
has the same eigenvalue $e^{i2\pi\tilde\alpha}$ in both
$\rho_{\tilde\alpha},\rho_{\tilde\alpha'}$,
by (\ref{irrepsn=1})$_2$ with $z=2\pi$
and the periodicity  $\psi(x\!+\!2\pi)=\psi(x)$.
In fact the inequivalent irreps of ${\rm Y}$ are parametrized by
$\tilde\alpha\in\mathbb{R}/\mathbb{Z}$
and are defined by (\ref{irrepsn=1})  up to  equivalences
(see \cite{Lou63,Mac63} or e.g. the more recent \cite{Tan93,Kas}).
$\{\frac {e^{ilx}}{\sqrt{2\pi}}
\}_{l\!\in\!\mathbb{Z}}
\subset C^\infty(S^1)$
is an orthonormal basis of $\L^2(S^1)$ consisting
of eigenvectors of $p$: \
$\rho_{\tilde\alpha}(p)e^{ilx}=(l\!+\!\tilde\alpha)e^{ilx}$. 

\subsubsection*{Irreducible unitary representations of \ $Y$ \ for \ $n=2$,
\ \ $\tilde\bA\neq 0$}

It is  $r=1$ and $[p_1,p_2]=i2\tilde b $, where
$\tilde\bA=\left(\!\ba{cc}\! 0\! & \!-\tilde b\! \\ 
\tilde b\! & \!0 \!\ea\!\right)$ and  $\tilde \nu\!=\!2\pi \tilde b\in\mathbb{N}$.
The decomposition (\ref{decoY}) becomes
$Y= M\, H_3$ where
$H_3\sim G$ and
$M$ is generated by $m_1,m_2$ fulfillig
$$
m_1m_2=e^{\frac {i\pi}{\tilde \nu}}m_2m_1.
$$
By the Von Neumann theorem
all irreps of $H_3$ are equivalent to
the Schr\"odinger representation on $\L^2(\mathbb{R})$.
On the other hand, the eigenvalues of the unitary casimirs
$\zeta_a$ ($a\!=\!1,2$) of $M$
necessarily have the form $e^{i2\pi\tilde\alpha_a}$,
$\tilde\alpha\in\mathbb{R}^2$;
hence the pairs $(e^{i2\pi\tilde\alpha_1},e^{i2\pi\tilde\alpha_2})$,
or equivalently the $\tilde\alpha\in\mathbb{R}^2/\mathbb{Z}^2$, identify
the classes of inequivalent irreps 
$(\rho_{\tilde\alpha},\H_{\tilde\alpha})$ 
of $M$, and therefore  of $Y$.
An irrep $(\rho_{\tilde\alpha},\X_{\tilde\alpha})$ 
of $\Oo$ is obtained on the subspace $\X_{\tilde\alpha}:=\H_{\tilde\alpha}\cap 
C^\infty(\mathbb{R}^2)$. The $p_a$ are essentially self-adjoint.
 When $\tilde\alpha=0$ $m_1,m_2$
are sometimes called `clock' and `shift'. 

From $\zeta_a=(m_a)^{2\tilde \nu}$ and (\ref{comrelm})
it also follows that each
$m_a$ has eigenvalues $e^{i \frac {\pi }{\tilde \nu}(\tilde\alpha_a+k)}$,
$k=0,1,...,2\tilde \nu\!-\!1$. The decomposition (\ref{direcstsumn}) takes the form
$\H_{\tilde \alpha}=\bigoplus_{k=0}^{2\tilde\nu-1} \H_{\tilde \alpha}^k$ 
(resp. $\X_{\tilde \alpha}=\bigoplus_{k=0}^{2\tilde\nu-1} \X_{\tilde \alpha}^k$),
and it is easy to check that
 $\H_{\tilde \alpha}^k$  (resp. $\X_{\tilde \alpha}^k$) is the eigenspace
of $m_1$ with eigenvalue $e^{i \frac {\pi }{\tilde \nu}(\tilde\alpha_1+k)}$.
$m_1$ and the whole $G\sim H_3$ map each $\H_{\tilde \alpha}^k$  
(resp. $\X_{\tilde \alpha}^k$)  into itself; only $m_2$ maps it outside, into 
$\H_{\tilde \alpha}^{k+1}$  (resp. $\X_{\tilde \alpha}^{k+1} $). [We identify
$\H_{\tilde \alpha}^{k+2\tilde \nu}=\H_{\tilde \alpha}^k$,  
$\X_{\tilde \alpha}^{k+2\tilde \nu}=\X_{\tilde \alpha}^k$].
Setting
$$
a\!:=\!\frac {p_1+i
p_2}{\sqrt{4\tilde b}},\qquad
\qquad a^*\!:=\!\frac {p_1-i
p_2}{\sqrt{4\tilde b}},\qquad
\qquad {\bf n}\!:=\!a^*a,
$$
we find $[a,a^*]=\1$.
An orthonormal basis   of the carrier
 spaces $\X_{\tilde\alpha},\H_{\tilde\alpha}$ 
consists of
eigenvectors \ $\psi_{{\rm n},k}$
\ (${\rm n}\in \mathbb{N}_0$, \ $k=0,1,...,2\tilde \nu\!-\!1$) \
of the complete set of commuting observables
$\{m_1,{\bf n}\}$:
\be
\ba{l}
\rho_{\tilde\alpha}(m_1)\psi_{{\rm n},k}=
e^{i \frac {\pi }{\tilde \nu}(\tilde\alpha_1+k)}\, \psi_{{\rm n},k},
\qquad\qquad\rho_{\tilde\alpha}({\bf n})\,\psi_{{\rm n},k}={\rm n}\, \psi_{{\rm n},k}.
\ea
\ee
Note that \ $a\, \psi_{0,k}=0$. The  \ $\psi_{{\rm n},k}$ \ with
\ ${\rm n}\in \mathbb{N}_0$ \ and fixed $k$ make up an orthonormal basis of \
$\X_{\tilde \alpha}^k,\H_{\tilde \alpha}^k$.\ 
Clearly the $\psi_{{\rm n},k}$ are also eigenvectors of the `Bochner Laplacian'
with constant magnetic field, $\Ha_{\tilde\nu}$:
\be
\ba{l}
\Ha_{\tilde\nu}\!:=\!\frac 12[(p_1)^2\!+\!(p_2)^2]=
\frac{\tilde \nu}{\pi}({\bf n}\!+\!\frac 12),\qquad\quad
\rho_{\tilde\alpha}(\Ha)\,\psi_{{\rm n},k}=E_{{\rm n}}\psi_{{\rm n}k},\qquad
 E_{{\rm n}}=
\frac{\tilde \nu}{\pi}\left({\rm n}\!+\!\frac 12\right).
\ea
\ee
The degeneracy of each ``energy level'' ${\rm n}$ is thus $2\tilde \nu$;
it  is related to the commutation relation
$[\Ha_{\tilde\nu},M]=0$, which in the usual treatment (see e.g. \cite{Zak,Tan})
is used as a condition defining the magnetic translation group $M$, whereas
here follows from the more general one $[U\g,M]=0$.
This finite degeneracy is well-known and in contrast with
the infinite degeneracy of Landau levels in
the absence of quasiperiodicity.
Choosing the gauge (\ref{quasiper}) with
$\beta$ as in (\ref{lowertriang}), the quasiperiodicity factor and
the decomposition (\ref{solutionsn}) of $\X_{\tilde\alpha},\H_{\tilde\alpha}$  take the form
\be
V(x)=e^{-i2\tilde \nu l_2 x^1},\qquad \qquad \psi(x)=
\sum\limits_{k=0}^{2\tilde\nu-1}\,\sum\limits_{l\in \mathbb{Z}}e^{i(k+2\tilde\nu l)x^1}
\psi_k(x^2\!+\!2\pi l); 
\ee
the subspaces \ $\X_{\tilde \alpha}^k,\H_{\tilde \alpha}^k$
\ are characterized by $\psi_{k'}\equiv 0$ for
$k'\in K\setminus\{k\}$. The formulae
\be
\ba{ll}
p_1\!=\!-i\partial_1\!+\!2\tilde  b x^2\!+\! \tilde  \alpha_1, \qquad &
p_2=-i\partial_2\!+\! \tilde \alpha_2,\\[8pt]
a=\frac{
\partial_2\!-\!i\partial_1\!+\!2
\tilde bx^2\!+\!\tilde \alpha_1\!+\!i
\tilde \alpha_2}{\sqrt{4\tilde b}},
\qquad &a^*=\frac{-
\partial_2\!-\!i\partial_1\!+\!2
\tilde bx^2\!+\!\tilde \alpha_1\!-\!i
\tilde \alpha_2}{\sqrt{4\tilde b}},\\[10pt]
m_1=e^{\frac {\pi}{\tilde \nu}(i\tilde\alpha_1+\partial_1)},
\qquad &m_2=e^{ix^1+\frac {\pi}{\tilde \nu}(i\tilde\alpha_2+\partial_2)},\\[8pt]
\zeta_1=e^{2\pi(i\tilde\alpha_1+\partial_1)},
\qquad &\zeta_2=e^{i2(\tilde\nu x^1+\pi\tilde\alpha_2)+2\pi\partial_2},\\[8pt]
\psi_{0,0}(x;\tilde\alpha,\tilde \nu) =N
\sum\limits_{l\in \mathbb{Z}}e^{i2\tilde \nu lx^1}e^{-\frac{\pi} {2\tilde \nu}
\left[\frac {\tilde \nu}{\pi} (x^2+2\pi l)+
\tilde \alpha_1+i\tilde\alpha_2\right]^2} |q\ra,\qquad
&\psi_{{\rm n},k}=\frac {(a^*)^{\rm n}}{\sqrt{{\rm n}!}} 
(m_2)^k\psi_{0,0},
\\[8pt]
a\psi_{{\rm n},k}=\sqrt{{\rm n}}\, \psi_{{\rm n}\!-\!1,k},\qquad
& a^*\psi_{{\rm n},k}=\sqrt{{\rm n}\!+\!1}\, \psi_{{\rm n}\!+\!1,k}
\ea                 \label{orthbasis2}
\ee
give the explicit representation of $\rho_{\tilde\alpha}$ and of the
above basis in this gauge; 
one easily determines the normalization constant to be
$|N|=\frac{\tilde\nu^{\frac 14}}{\sqrt{2}\pi}
e^{-\frac{\pi(\tilde \alpha_2)^2}{2\tilde\nu}}$.
$\psi_{0,0}$ is cyclic. 
Introducing the complex variables  $z=\frac {\tilde \nu}{\pi}(x^1\!\!+\!i x^2)$,
 $\bar z=\frac {\tilde \nu}{\pi}(x^1\!\!-\!i x^2)$  \ we find 
\be
\ba{l}
a=\sqrt{\frac{\pi}{2\tilde \nu}} \left[\frac i2(\bar z\!-\! z)\!+\!
i\frac {2\tilde \nu}{\pi}\partial_{\bar z}\!+\!\tilde \alpha_1\!+\!i\tilde \alpha_2\right],
\qquad\quad a^*=\sqrt{\frac{\pi}{2\tilde \nu}} \left[\frac i2(\bar z\!-\! z)\!+\!
i\frac {2\tilde \nu}{\pi}\partial_z\!+\!\tilde \alpha_1\!-\!i\tilde \alpha_2\right],\\[14pt]
\psi_{0,k}\big(x;\tilde \alpha,\tilde \nu\big)=
g(x^2)\: e^{\frac{\pi} {\tilde \nu}[ik(z+i\tilde\alpha_1\!-\!\tilde \alpha_2)-
\frac 12 k^2]}\: 
\vartheta\!\left[z\!+\!i\tilde \alpha_1\!+\!ik\!-\!\tilde \alpha_2,
i2\tilde \nu\right]|q\ra.
\ea                     \label{orthbasis2'}
\ee
where we have set \ $g(x^2)\!:=\!Ne^{-\frac{\pi} {2\tilde \nu}
\left[\frac {\tilde \nu}{\pi} x^2+
\tilde \alpha_1+i\tilde\alpha_2\right]^2}$ \ and used Jacobi  Theta function
\be
\vartheta(z,\! \tau)\!:=\!
\sum\limits_{l\in \mathbb{Z}}\! e^{i\pi l(2z+l\tau)}. \label{JTheta}
\ee
Hence, up to the gaussian factor
$g(x^2)$ the 
$\psi_{0,k}$ 
are
{\it analytic} (actually {\it entire}) functions of 
$z$\footnote{
Eq. (\ref{orthbasis2'}) is verified by direct inspection for
$({\rm n},k)\!=\!(0,0)$, using the relation
$m_2f(x^1,x^2)\!=\!e^{ix^1+i\frac {\pi}{\tilde \nu}\tilde\alpha_2}
f\!\!\left(x^1,x^2\!+\!\frac{\pi} {\tilde \nu}\!\right)$
[valid for all
$f\in C^\infty(\mathbb{R})$] and (\ref{orthbasis2}).}.
Applying the transformation
\be
\psi\mapsto \psi':= g^{-1}\psi,\qquad\qquad
V(l,x)\mapsto V'(l,x)=g^{-1}(x\!+\!2\pi l)V(l,x)g(x),
\ee
to the $\psi_{0,k}$ one obtains $\psi_{0,k}'$
depending on $x^1,x^2$ only through the
complex variables $z$, w.r.t. which they are holomorphic.
The hermitean structure (\ref{hermstr})
 and the action of $\Oo,\Y$
are resp. transformed as follows: 
and
\be
dx^1dx^2\mapsto  dx^1dx^2 |g|^2,\qquad\qquad
w\mapsto w':=  g^{-1}\, w\, g,\quad \forall w\in\Oo,\Y.
\ee
In particular, \ $a'=i\sqrt{\frac{2\tilde \nu}{\pi}} \partial_{\bar z}$, \ 
$a^*{}'=\sqrt{\frac{\pi}{2\tilde \nu}}i(\bar z\!-\! z)\!-\!i
\sqrt{\frac{2\tilde \nu}{\pi}} \partial_z$; \  moreover, if $\tilde \alpha=0$ \  one finds
\be
m_1'=e^{\frac {\pi}{\tilde \nu}\partial_1},
\qquad m_2'=e^{\frac {\pi}{\tilde \nu}\left(iz-\frac 12+\partial_2\right)},\qquad
\psi_{0,k}'=e^{\frac{\pi} {\tilde \nu}[ikz-\frac 12 k^2]}\: 
\vartheta\!\left[z\!+\!ik\!,i2\tilde \nu\right]|q\ra.
\ee
This shows the link with
the holomorphic framework (see e.g. \cite{Mum83,BirLan04})
on (the universal cover of) a complex torus with parameter  \  $\tau=i2\tilde \nu$: \ 
$V'$ \  is called an {\it automorphy factor} and takes values
in \  $\b{C}\setminus\{0\}$.


\subsubsection*{Irreducible unitary representations of $Y$ for
general $n$}
We are now ready for tori of general dimension $n$. Using the
transformation (\ref{modular}), the decomposition (\ref{decoY}),
the results for $n=1,2$,
the equivalence (by Von Neumann theorem) of all irreps of
$H_{2r\!+\!1}$ to
the Schr\"odinger representation on $\L^2(\mathbb{R}^r)$,
we find the following

\begin{prop}
The sets of joint eigenvalues $\zeta_a=e^{i2\pi\tilde\alpha_a}$ of the unitary Casimirs, 
or equivalently the $\tilde\alpha\in\mathbb{T}^n_J:=\mathbb{R}^n/\mathbb{Z}^n$,
identify up to unitary equivalences the irreducible unitary representations (irreps)
$(\rho_{\tilde\alpha},\H_{\tilde\alpha})$ of $Y$ and $\Y$. 
An irrep $(\rho_{\tilde\alpha},\X_{\tilde\alpha})$ 
of $\Oo$ is obtained by restriction
on the subspace $\X_{\tilde\alpha}:=\H_{\tilde\alpha}\cap 
C^\infty(\mathbb{R}^n)$.  The $p_a$ are essentially self-adjoint.

For $j=1,...,r$, each $m_j$ has eigenvalues 
$e^{i \frac {\pi }{\tilde \nu}(\tilde\alpha_j+k_j)}$,
$k_j=0,1,...,2\tilde \nu_j\!-\!1$. 
The decomposition (\ref{direcstsumn}) takes the form
$\H_{\tilde \alpha}=\bigoplus_{k\in K} \H_{\tilde \alpha}^k$ 
(resp. $\X_{\tilde \alpha}=\bigoplus_{k\in K} \X_{\tilde \alpha}^k$),
where $\H_{\tilde \alpha}^k$  (resp. $\X_{\tilde \alpha}^k$) is the eigenspace
with eigenvalues 
$e^{i \frac {\pi }{\tilde \nu}(\tilde\alpha_1+k_1)},...,
e^{i \frac {\pi }{\tilde \nu_r}(\tilde\alpha_r+k_r)}$ of \ $m_1,...,m_r$.
The latter operators and the whole $G$ (resp. $U\g$)
map each $\H_{\tilde \alpha}^k$  (resp. $\X_{\tilde \alpha}^k$) 
into itself. The $m_{r\!+\!j}$, map it  into 
$\H_{\tilde \alpha}^{k+e_j}$  (resp. $\X_{\tilde \alpha}^{k+e_j} $). [We identify
$\H_{\tilde \alpha}^{k+2\tilde \nu_je_j}=\H_{\tilde \alpha}^k$,  
$\X_{\tilde \alpha}^{k+2\tilde \nu_je_j}=\X_{\tilde \alpha}^k$].
Setting $r:=\frac 12\mbox{rank}(\tilde \bA )\le\frac 12 n$ and
performing first a transformation
(\ref{modular}) leading to $(\nu_1,...,\tilde\nu_r)\in\mathbb{N}^r$
with $\nu_{j\!+\!1}/\nu_j\in\mathbb{N}$, and then defining
\be
\tilde b_j\!:=\!\frac{\tilde\nu_j}{2\pi},\qquad
a_j\!:=\!\frac {p_j+i
p_{r\!+\!j}}{\sqrt{2\tilde b_j}},\qquad
a_j^*\!:=\!\frac {p_j-i
p_{r\!+\!j}}{\sqrt{2\tilde b_j}},\qquad
{\bf n}_j\!:=\!a_j^*a_j
\ee
for $j=1,...,r$ we find that $[a_i,a_j^*]=\1\delta_{ij}$.
An orthonormal basis $\B$  of the carrier Hilbert space $\H_{\tilde\alpha}$
consists of joint eigenvectors  $\psi_{{\rm n},k,l}$ of
the complete subset of commuting observables
$\{{\bf n}_1,...,{\bf n}_r,m_1 ,...,m_{r},p_{2r\!+\!1},...,p_n \}
\subset \Oo$:
\be
\ba{l}
{\rm n}\in\mathbb{N}_0^r,\qquad\qquad k
\in K:=\{0,...,2\tilde\nu_1\!-\!1\}\times...\times
\{0,...,2\tilde\nu_r\!-\!1\},
\qquad\qquad l\in\mathbb{Z}^{n\!-\!2r} \\[8pt]
{\bf n}_j\,\psi_{{\rm n},k,l}={\rm n}_j\, \psi_{{\rm n},k,l},
\qquad m_{j} \,\psi_{{\rm n},k,l}=e^{ \frac {i\pi }{\tilde \nu_j}
\left[\tilde\alpha_{j}+k_j\right]}\, \psi_{{\rm n},k,l},
\qquad p_a\,\psi_{{\rm n},k,l}=(l_a\!\!+\!\tilde\alpha_a\!)\, \psi_{{\rm n},k,l},
\ea
\ee
$a\!>\!2r$. It is $a_j \psi_{0,k,l}=0$. An orthonormal basis
of the subspace $\H_{\tilde\alpha}^k,\X_{\tilde\alpha}^k$  consists of
the $\psi_{{\rm n},k,l}$ with that $k$ only.

More explicitly, in the gauge (\ref{quasiper}) with $\beta$ as in (\ref{lowertriang})
we obtain
\be
\ba{l}
p_j\!=\!-\!i\partial_j\!+\!2\tilde  b_j x^{r\!+\!j}\!\!+\! \tilde\alpha_j,
\qquad\qquad \qquad\quad
p_a\!=\!-i\partial_a\!+\!\tilde\alpha_a,\qquad a=r\!+\!1,...,n,\\[8pt]
a_j=\frac{
\partial_{r\!+\!j}\!-\!i\partial_j \!+\!2
\tilde b_jx^{r\!+\!j}\!+\!\tilde\alpha_j\!+\!i
\tilde\alpha_{r\!+\!j}}{\sqrt{4\tilde b_j}},
\qquad \quad\qquad\quad a_j^*=\frac{
-\partial_{r\!+\!j}\!-\!i\partial_j\!+\!2
\tilde b_jx^{r\!+\!j}\!+\!\tilde\alpha_j\!-\!i
\tilde\alpha_{r\!+\!j}}{\sqrt{4\tilde b_j}}, \\[10pt]
m_j=e^{\frac {\pi}{\tilde \nu_j}(i\tilde\alpha_j+\partial_j)},
\qquad\qquad\qquad\qquad\quad \:  m_{r\!+\!j}=e^{ix^j+\frac {\pi}{\tilde \nu_j}
(i\tilde\alpha_{r\!+\!j}+\partial_{r\!+\!j})},\\[8pt]
\zeta_j=e^{2\pi(i\tilde\alpha_j+\partial_j)},
\qquad\qquad\qquad\qquad\qquad \zeta_{r\!+\!j}=e^{i2\tilde\nu_j x^j\!+\!2\pi(
i\tilde\alpha_{r\!+\!j}+\partial_{r\!+\!j})},\\[8pt]
\psi_{0,0,0}(x;\tilde\alpha,\tilde \nu) =N'\,
\sum\limits_{l\in \mathbb{Z}^r}\exp\left\{\sum\limits_{j=1}^r\left[i2\tilde\nu_jl_jx^j
-\frac {\pi}{2\tilde \nu_j}
 \left(\frac{\tilde \nu_j}{\pi} x^{r\!+\!j}\!+\!2\tilde\nu_jl_j\!+\!\tilde\alpha_j\!+\!i
\tilde\alpha_{r\!+\!j}\right)^2\right]\right\} |q\ra, \\[12pt]
\psi_{{\rm n},k,l}=\left[\prod\limits_{a=2r\!+\!1}^ne^{il_ax^a}\right]
\left[\prod\limits_{j=1}^r \frac {(a_j^*)^{{\rm n}_j}}
{\sqrt{{\rm n}_j!}} (m_{r\!+\!j})^{k_j}
\right]\,\psi_{0,0,0}\\[10pt]
a_j\psi_{{\rm n},k,l}=\sqrt{{\rm n}_j}\, \psi_{{\rm n}\!-\!e_j,k},\qquad
\qquad\qquad\qquad 
a^*\psi_{{\rm n},k,l}=\sqrt{{\rm n}_j\!+\!1}\, \psi_{{\rm n}\!+\!e_j,k,l};
\ea                 \label{orthbasisn}
\ee
the normalization constant is given by
$|N'|=\frac 1{(2\pi)^{n/2}} \prod_{j=1}^r
\left(\sqrt{2}\tilde\nu_j^{\frac 14}\right)
\exp\left[-\frac{\pi(\tilde \alpha_{r\!+\!j})^2}{2\tilde\nu_j}\right]$.
By gauge transformations (\ref{gaugetr}) one explicitly obtains the other
equivalent representations.
\end{prop}

{\bf Bibliographical note}.
$M:=M^1...M^r$ defined in (\ref{decoY}) is the
``group of magnetic translations'', in the sense
of Zak \cite{Zak}; it is a subgroup of the commutant
$\tilde M$ of $G$ within $Y$\footnote{The whole
commutant (centralizer)  $\tilde M$ of $G$ within $Y$ is
the subgroup
$$
\ba{l}
\tilde M=M M',\qquad \qquad
M'\!:=\left\{\!\exp\left[ih^0\!+\!i\sum_{a=2r\!+\!1}^np_a
z^a\right]  \:\: |
\:\:(h^0\!,\!z^{2r\!+\!1}\!,...,\!z^n)\in\mathbb{R}^{n\!-\!2r\!+\!1} \right\}
\ea
$$
Equivalently, $[m,U\g]=0$, i.e. $[m,p_a]=0$, for all
$m\in \tilde M$.
}. 
In the literature the torus $\mathbb{T}^n_J$, to which $\tilde\alpha$ belongs,
is sometimes called `Brillouin zone', or  `Jacobi torus'.

\medskip
Clearly, $\psi_{{\rm n},k,l}$ are also eigenvectors of the `Bochner Laplacian'
with constant $B$, $\Ha_{\bA}$:
\be
\ba{l}
\Ha_{\bA}:=\sum\limits_{a=1}^n(p_a)^2=\sum\limits_{j=1}^r
\frac{\tilde \nu_j}{\pi}(a^*_ja_j\!+\!\frac 12)
\!+\!\sum\limits_{a=2r\!+\!1}^n\!p_a^2,\\
\Ha\,\psi_{{\rm n},k,l}=E_{{\rm n},l}\psi_{{\rm n},k,l},\qquad\qquad
\quad E_{{\rm n},l}=\sum\limits_{j=1}^r
\frac{\tilde \nu_j}{\pi}({\rm n}_j\!+\!\frac 12)\!+\!
\sum\limits_{a=2r\!+\!1}^n\!(l_a\!\!+\!\tilde\alpha_a)^2.
\ea
\ee
Again, the finite degeneracy $\prod_{j=1}^r(2\tilde\nu_j)$ of the energy level
$E_{{\rm n},l}$  is related to the commutation relation
$[\Ha_{\bA},M]=0$, which in the usual treatment (see e.g. \cite{Zak,Tan})
is used as a condition defining the magnetic translation group $M$, whereas
here follows from the more general one $[U\g,M]=0$.

The $\psi_{0,k,0}$ are related to Jacobi Theta function
$\vartheta$ and Riemann  Theta function
$$
\theta(z, \tau):=
\sum\limits_{l\in \mathbb{Z}^r}e^{i\pi (2l^tz+l^t\tau l)},\qquad
z\!\equiv\!(z^1,...,z^r)\!\in\!\b{C}^r, \quad\tau\!\in\!\mathbb{H}_r\!:=\!\{\tau\!\in\!
 M_r(\b{C})\:\:|\:\: \tau^t\!=\!\tau,\,\Im(\tau)\!>\!0\}
$$
($\mathbb{H}_r$ is {\it Siegel upper half space}) by
\be
\ba{l}
\psi_{0,k,0}(x;\tilde \alpha,\tilde \nu)\:=
G(x)\left[\prod\limits_{j=1}^re^{\frac{\pi} {\tilde \nu_j}
\left(ik_j z^j_{k,\alpha}+\frac 12 k_j^2\right)} \vartheta\!\left(\!z^j_{k,\alpha},
i2\tilde \nu_j\!\right)\!\right]|q\ra \\[14pt]
\qquad\qquad\qquad = G(x) \: \left[\prod\limits_{j=1}^re^{\frac{\pi} {\tilde \nu_j}
\left(ik_j z^j_{k,\alpha}+\frac 12 k_j^2\right)} \!\right]\theta\!\left(z_{k,\alpha}, i2\tilde \nu\right)|q\ra\\[14pt]
G(x)\!:=\!N'\,\exp\!\left[-\sum\limits_{j=1}^r\frac{\pi}{2\tilde \nu_j}
(\frac{\tilde \nu_j}{\pi}x^{r\!+\!j}\!+\!\alpha_{r\!+\!j}\!+\!i\alpha_j
)^2\right] ,\\[14pt]
 z^j_{k,\alpha}\!:=\!z^j\!+\!ik_j\!-\!\tilde 
\alpha_{r\!+\!j}\!+\!i\alpha_j,\qquad\quad
z^j\!:=\!\frac {\tilde \nu_j}{\pi}(x^j\!+\!i x^{r\!+\!j}),\qquad\quad 
\tilde \nu\!:=\!\mbox{diag}(\tilde \nu_1,...,\tilde \nu_r).
\ea                               \label{RTheta}
\ee
Hence, up to the gaussian factor $G(x)$ the  $\psi_{0,k,0}$ are {\it analytic} 
(actually {\it entire}) functions of the $z^j$, $j=1,...,r$. 

\medskip
The above results, propaedeutical for the torus,
are useful also for the physics
of a scalar charged particle on $\mathbb{R}^n$.
The Hilbert space of states is \cite{AscOveSei94,Gru00}
 the direct integral
over $\tilde\alpha\in\mathbb{T}^n_J$ of the Hilbert spaces
$\H^V_{\tilde\alpha}$ of the inequivalent representations of $\OQ$
parametrized by $\tilde\alpha\in\mathbb{T}^n_J$. In the presence of
a periodic scalar potential ${\rm V}$ and a periodic magnetic field
$B$ with fixed fluxes (\ref{quantizedB})
 the Hamiltonian $\Ha$ of (\ref{1Schr})
belongs to $\OQ$, and the corresponding evolution is such that
$\psi(t_0)\in\H^V_{\tilde\alpha}$
implies $\psi(t)\in\H^V_{\tilde\alpha}$ for all $t$.

\section{Mapping \ $\X^V\stackrel{\sim}{\rightarrow} \Gamma(\mathbb{T}^n\!,\!E)$}
\label{VtoE}

The 1-particle wavefunction describing a charged scalar particle
on an orientable Riemannian manifold $X$ in the presence of a  magnetic
field 2-form $B=B_{ab}dx^adx^b$ is
a $\L^2$-section of a hermitean line bundle
$E\stackrel{\pi}{\mapsto}X$ with connection having field strength
$B$ (see e.g. \cite{WuYan75,Kos70,Woo92});
if $X$ is compact the fluxes of $B$ are necessarily quantized.
We now show that if $X\!=\!\mathbb{T}^n\!=\!\mathbb{R}^n/2\pi\mathbb{Z}^n$
then also $E$, beside $X$, can be realized as a quotient,
and its $\L^2$-sections are determined
by the quasiperiodic wavefunctions
$\psi(x)$ on $\mathbb{R}^n$ introduced in the earlier sections.
We partly mimic arguments used for complex tori
(see e.g. appendix B in \cite{BirLan04}).

The formula
$T_l:x\in\mathbb{R}^n\mapsto x\!+\!2\pi l\in\mathbb{R}^n$  ($l\in\mathbb{Z}^n$)
defines an action of the  abelian group $\mathbb{Z}^n$ on $\mathbb{R}^n$.
The action  is free, in that $T_l(x)=x$ for
some $x$ implies that $T_l=\id=T_0$, and properly discontinuous, in that
the inverse image of any compact subset is compact.
Clearly $T_l\circ T_{l'}=T_{l\!+\!l'}$
and  $T_l^{-1}=T_{-l}$.
Introducing in $\mathbb{R}^n$ the equivalence relation
$x\sim x'$ iff $x'=T_l(x)$ for some $l\in\mathbb{Z}^n$, the definition of
the torus $\mathbb{T}^n$ as a quotient $\mathbb{R}^n/2\pi\mathbb{Z}^n$ amounts to
\be
\mathbb{T}^n=\mathbb{R}^n/\!\sim\,;           \label{defL}
\ee
in other words, an element of $\mathbb{T}^n$ is an equivalence class
$[x]=\{T_l(x),\:\: l\in\mathbb{Z}^n \}$. The universal cover
map is defined as  $P:x\in\mathbb{R}^n\mapsto [x]\in\mathbb{T}^n$.
The fundamental cells $C^y_{a_1...a_k}$ defined in
(\ref{fundamentalcells}) are mapped
by $P$ onto fundamental $k$-cycles
$\tilde C^{[y]}_{a_1...a_k}\subset \mathbb{T}^n$ through
$[y]$. The $dx^a$ generate the exterior algebra $\bigwedge^*$ not only
on $\mathbb{R}^n$ but also on $\mathbb{T}^n$, and by multiplication
by $f\in\X$ and linearity the algebra $\Omega^*(\mathbb{T}^n)$
of differential forms on $\mathbb{T}^n$. We can consider $\Omega^*(\mathbb{T}^n)$
as a subspace of $\Omega^*(\mathbb{R}^n)$, and the exterior derivative $d$
on $\Omega^*(\mathbb{T}^n)$  as the restriction of $d$
on $\Omega^*(\mathbb{R}^n)$.

Similarly,
given a smooth phase factor $V:\mathbb{Z}^n\times \mathbb{R}^n\mapsto U(1)$
fulfilling (\ref{U1cocycle}), the formula
\be
\chi^V_l:(x,c)\in\mathbb{R}^n\times\b{C}\mapsto \Big(x\!+\!2\pi l\,,\,V(l,x)\,c\Big),\qquad\qquad  l\in\mathbb{Z}^n
\ee
defines an action of the abelian group $\mathbb{Z}^n$ on $\mathbb{R}^n\times\b{C}$.
The action is free, in that $\chi_l^V[(x,c)]=(x,c)$ for
some $(x,c)$ implies that $\chi_l^V=\id=\chi^V_0$, and again  properly discontinuous.
By (\ref{U1cocycle}), $\chi^V_l\circ\chi^V_{l'}=\chi^V_{l\!+\!l'}$
and  $(\chi^V_l)^{-1}=\chi^V_{-l}$.
In $\mathbb{R}^n\times\b{C}$ we introduce an equivalence relation $\sim_V$
by setting
$(x,c)\sim_V (x',c')$ iff $(x',c')=\chi^V_l[(x,c)]$ for some $l\in\mathbb{Z}^n$;
we correspondingly define
\be
E=(\mathbb{R}^n\times\b{C})/\sim_V;           \label{defE}
\ee
in other words, an element of $E$ is an equivalence class
$[(x,c)]=\{\chi^V_l\big((x,c)\big),\:\: l\in\mathbb{Z}^n \}$.
The projection $\pi: E\mapsto \mathbb{T}^n$ is defined by
$\pi\big([(x,c)]\big)=[x]$.
$E$ is trivial (i.e. $E=\mathbb{T}^n\times\b{C}$)
if $V$ is trivial [i.e. $V(l,x)\equiv 1$].

Given a smooth function $\psi: \mathbb{R}^n\mapsto\b{C}$ fulfilling
(\ref{quasiperiodicity}) we can define a smooth map
$$
\bpsi: [x]\in\mathbb{T}^n\mapsto\left[\!\Big(\!x,\psi(x)\!\Big)\!\right]
=\left\{\chi^V_l\!\left[\!\Big(\!x,\psi(x)\!\Big)\!\right],
\:\: l\!\in\!\mathbb{Z}^n \right\}
\stackrel{(\ref{quasiperiodicity})}{=}\left\{
\Big(\!x\!+\!2\pi l,\psi(x\!+\!2\pi l)\!\Big),
 \:\: l\!\in\!\mathbb{Z}^n \right\}\in E,
$$
i.e. a (global) section $\bpsi\in\Gamma(\mathbb{T}^n,E)$.
The correspondence $\phi:\psi\in\X^V\mapsto\bpsi\in\Gamma(\mathbb{T}^n,E)$
is one-to-one. Given $\bpsi,\bpsi'\in\Gamma(\mathbb{T}^n,E)$, then
$\overline{\bpsi'}\bpsi([x]):=\overline{\psi'(x)}\psi(x)$
defines a function $\overline{\bpsi'}\bpsi\in\X$, and
\be
(\bpsi',\bpsi)_\Gamma:=\int_{\mathbb{T}^n}\!\!\!d^nx\:\,
\overline{\bpsi'}\,\bpsi=\int_{C^y_{1...n}}\!\!\!\!\!\!d^nx\:\,
\overline{\psi'(x)}\,\psi(x)\stackrel{(\ref{hermstr})}{=} (\psi',\psi)
\label{hermstr'}
\ee
a hermitean structure $(~,~)_\Gamma$
equal to that of $\X^V$ and making $\Gamma(\mathbb{T}^n\!,\!E)$
a hermitean line bundle.
A compatible covariant derivative is defined by
\be
\bnabla:=\phi\circ\nabla\circ \phi^{-1},
\qquad\qquad \bnabla: \Omega^p({\mathbb{T}^n})\otimes_{\X}\Gamma(\mathbb{T}^n\!,\!E)
\mapsto \Omega^{p\!+\!1}({\mathbb{T}^n})\otimes_{\X}\Gamma(\mathbb{T}^n\!,\!E).
\ee
The curvature map
$\bnabla^2: \Gamma(\mathbb{T}^n\!,\!E)\mapsto \Omega^2({\mathbb{T}^n})
\otimes_{\X}\Gamma(\mathbb{T}^n\!,\!E)$
determines the field strength 2-form $B$ on $\mathbb{T}^n$ through
the formula  $\bnabla^2\bpsi=-2iqB\otimes_{\X}\bpsi$.
Relations (\ref{fluxes}-\ref{fluxesm}) become
\be
\phi_{ab}=\int_{\tilde C_{ab}^{[y]}}\!\!\!  B =2\pi\nu_{ab},
\qquad\qquad\int_{\tilde C_{a_1...a_{2m}}^{[y]}}\!\!\! \!\!\! B^m=
(2\pi)^m\nu_{[a_1a_2}\nu_{a_{2m\!-\!1}a_{2m}]}
\ee
(the result is independent of $[y]$).
In particular $\phi_{ab}$ is the flux of $B$  through a
2-cycle $\tilde C^{[y]}_{ab}$; therefore $E$
is characterized by the Chern numbers (\ref{quantizedB}).
The Hilbert space completion of  $\Gamma(\mathbb{T}^n\!,\!E)$ with
hermitean structure (\ref{hermstr'}) is isomorphic to $\H^V$.
The actions of $\Oo,\g,Y,G$ are lifted from $\X^V,\H^V$ to
$\Gamma(\mathbb{T}^n\!,\!E)$ and the completion of the latter by replacing
$\rho(o)\to \phi\circ\rho(o)\circ \phi^{-1}$, where
$o\in \Oo,\g,Y,G$ respectively. For instance we find
\be
[g_{\tilde z}\bpsi]([x]):=\left[\!\Big(\!x,[g_{\tilde z}\psi]
(x)\!\Big)\!\right].                  \label{GQaction0'}
\ee

Given any smooth gauge transformation $U(x)$ on $\mathbb{R}^n$,
the replacements
$\big(V,\psi,\nabla\big)\to \big(V^U,\psi^U,U\nabla U^{-1}\big)$
[see (\ref{gaugetr})] result into  a new hermitean line bundle
$E^U=(\mathbb{R}^n\times\b{C})/\sim_{V^U}$ isomorphic to $E$,
a section $\bpsi^U\in\Gamma(\mathbb{T}^n,E^U)$ isomorphic to
$\bpsi\in\Gamma(\mathbb{T}^n,E)$ and a covariant derivative
$\bnabla^U$ on $\Gamma(\mathbb{T}^n,E^U)$ isomorphic to
$\bnabla$ on $\Gamma(\mathbb{T}^n,E)$; in other words,
they result into a gauge transformation on $\mathbb{T}^n$.
The $m$-th Chern class of $E$ is given by the gauge-invariant
\be
\mbox{Ch}_m= \left[\frac{B^m}{m!(2\pi)^m}\right]= \left[\frac{(\bbA)^m}{m!(2\pi)^m}\right],                \label{Chern}
\ee
where $[\omega]$ stands for the de Rham cohomology class
containing the closed $p$-form $\omega$.

\bigskip
The above data determine also trivializations of
$E,\Gamma(\mathbb{T}^n,E),\bnabla$ in a canonical way.
For each set $X_i$
of a (finite) open cover $\{X_i\}_{i\in\I}$ of $\mathbb{T}^n$ let
$W_i$ be a subset of $\mathbb{R}^n$ such that the restriction
$P_i\equiv P:W_i\mapsto X_i$ is invertible. For $u\in X_i$ let
\be
\bpsi_i(u):=\psi[P_i^{-1}(u)], \quad\qquad A_{ia}(u):=A_a[P_i^{-1}(u)],
\qquad\quad \nabla\!_i:=-id\!+\!qA_i.
                                    \label{deftrivialization}
\ee
As a consequence of (\ref{quasiperiodicity}) we find
in $X_i\cap X_j$\footnote{
The points  $x\!\in\! W_j$, $x'\!\in\! W_i$ such that $u=P_j x=P_i x'$
are related by $x'=x\!+\!2\pi l$, with some $l\in\mathbb{Z}^n$. One has just
to replace the arguments $l,x$ of $V$ in (\ref{quasiperiodicity})
resp. by $P_i^{-1}(u)\!-\!P_j^{-1}(u)$, $P_j^{-1}(u)$.}
\be
\ba{l}
\bpsi_i=t_{ij}\bpsi_j,\qquad \nabla\!_i=t_{ij}
\nabla\!_jt_{ji},\qquad\quad
t_{ij}(u)\! :=\! V\!\left\{\!\frac 1{2\pi}\!
\!\left[P_i^{-1}(u)\!-\!P_j^{-1}(u)\right],P_j^{-1}(u)\!\right\}
\ea                         \label{trivialization}
\ee
Condition (\ref{U1cocycle}) becomes\footnote{
The points  $x\!\in\! U_k$,
$x'\!\in\! W_j$, $x''\!\in\! W_i$ such that
$[x]=[x']=u=P_i x''=P_j x'=P_k x$
are related by $x'=x\!+\!2\pi l'$, $x''=x'\!+\!2\pi l$
with some $l,l'\in\mathbb{Z}^n$. One has to replace
 $x,x\!+\!2\pi l',l,l \!+\!l'$  in (\ref{U1cocycle})
resp. by $P_k^{-1}(u),P_j^{-1}(u),\left[P_i^{-1}(u)\!-\!P_j^{-1}(u)
\right]/2\pi, \left[P_i^{-1}(u)\!-\!P_k^{-1}(u)\right]/2\pi$,
and use the above definition of $t_{ij}$.
}
\be
t_{ik}=t_{ij}t_{jk},\qquad\qquad \quad \mbox{ in } X_i\cap X_j\cap X_k.
                                            \label{tfcocycle}
\ee
This can be interpreted as the ($\check{\rm C}$ech cohomology) cocycle
condition for the transition functions $t_{ij}$ of
the hermitean line $E$ defined in (\ref{defE}).
The set $\{(X_i,\bpsi_i,\nabla_i) \}_{i\in\I}$ defines a trivialization
of the section $\bpsi\in\Gamma(\mathbb{T}^n\!,\!E)$ of $E$ and of the
compatible covariant derivative $\bnabla$.
The hermitean structure
can be expressed in terms of the trivialization as
$(\bpsi',\bpsi)_\Gamma:=\sum_\tau\int_{X'_\tau}\!\! d^nx\:
\overline{\bpsi'_{i(\tau)}}\bpsi_{i(\tau)}$,
where $\{X'_\tau\}_{\tau\in T}$ is a partition of $\mathbb{T}^n$
such that for all
$\tau\in T$  it is $X'_\tau\subset X_{i(\tau)}$ for some $i(\tau)\in\I$.
The advantage of the definition (\ref{hermstr'}) is that
we don't have to bother about patch-dependent $\bpsi_i$.
Setting $U_i(u):=U[P_i^{-1}(u)]$ for $u\in X_i$,
the set $\{(X_i,U_i) \}_{i\in\I}$ defines the trivialization
of a gauge transformation:
\be
\bpsi_i\mapsto\bpsi_i^U=U_i\bpsi_i, \quad\qquad
t_{ij}^U=U_it_{ij}U^{-1}_j,
\qquad\quad \nabla\!_i\mapsto\nabla\!_i^U=U_i\nabla\!_iU_i^{-1};
\ee
if $U(x)$ is periodic the transformation
is {\it globally defined}, $U_i(u)=U_j(u)$, and $t^U_{ij}=t_{ij}$.

For fixed cover $\{X_i\}_{i\in\I}$, a  different choice $\{\tilde
W_i\}_{i\in\I}$ of the $\{W_i\}_{i\in\I}$ in the above construction
amounts to a gauge transformation $\{(X_i,\tilde
U_i)\}_{i\in\I}$\footnote{It must be $\tilde W_i=W_i\!+\!2\pi l_i$
for some  $l_i\in\mathbb{Z}^n$, whence $\tilde {\bpsi}_i(u)=\psi\big[\tilde
P_i^{-1}(u)\big]=\psi\big[P_i^{-1}(u)\!+\!2\pi l_i\big]=$
$V\big[l_i,P_i^{-1}(u))\big]\psi\big[P_i^{-1}(u)\big]=
\tilde U_i(u)\bpsi_i(u)$, where $\tilde
U_i(u)\!:=\!V\big[l_i,P_i^{-1}(u))\big]$.
},
in agreement with the fact that it leads to the same bundle $E$.

For any $[z]\in\mathbb{T}^n$ let
$T_{[z]}\!:\!\mathbb{T}^n\!\mapsto\! \mathbb{T}^n$ be the translation operator
$T_{[z]}[x]:=[x\!+\!z]$.
In terms of the trivialization
$\{(X_i,[g_{\tilde z}\bpsi]_i) \}_{i\in\I}$ (\ref{GQaction0'}) reads
\be
[g_{\tilde z}\bpsi]_i(u):=[g_{\tilde z}\psi][P_i^{-1}(u)]
\stackrel{(\ref{gauge0})}{=}e^{iq\left\{z^0+\alpha^t z+
[P_i^{-1}(u)]^t\!\bA z\!\right\}}\:  \bpsi_j\big(T_{[z]}u\big)       \label{GQaction'}
\ee
where $X_j$ is such that $T_zu\in X_j$\footnote{As
$P(x\!+\!z)=T_{[z]}u\in X_j$, then $x\!+\!z=P_j^{-1}\big(T_{[z]} u\big)$,
whereas $x=P_i^{-1}(u)\in X_i$; replacing these formulae in
(\ref{GQaction}) we obtain the second equality in (\ref{GQaction'}).
As a consistency check, 
it is straightforward to verify that
the conditions $[g_{\tilde z}\psi]_i=t_{ij}
[g_{\tilde z}\psi]_j$ are satisfied.
}, and the second equality holds  in the gauge (\ref{gauge0}) only.
Eq. (\ref{GQaction'}) could have been hardly guessed
without this lifting procedure, as it is {\it non-local}
(the result for the $i$-th component involves
other components).

\section{Conclusions}
\label{conclu}

Starting from the basic notion of {\it quasiperiodicity factors}
$V:\mathbb{Z}^n\times\mathbb{R}^n\mapsto U(1)$,
in the first three sections we have introduced in a systematic way
tools and general results regarding quantum mechanics of
a charged scalar particle on $\mathbb{R}^n$ in the presence
of a magnetic field and a scalar potential periodic under
discrete translations in a lattice $\Lambda$ of maximal rank.

In section \ref{VtoE} we have shown how these tools and results
can be reinterpreted and
re-used for the same theory on the torus $\mathbb{T}^n=\mathbb{R}^n/\Lambda$,
using the one-to-one correspondences between
smooth quasiperiodicity factors $V$
and hermitean line bundles $E\stackrel{\pi}{\mapsto}\mathbb{T}^n$,
pre-Hilbert spaces $\X^V$ of the type (\ref{quasiperiodicity}) \& (\ref{U1cocycle})
and pre-Hilbert spaces $\Gamma(\mathbb{T}^n\!,\!E)$ of smooth sections of $E$,
and between the covariant derivatives, algebras/groups of observables
acting on  $\X^V,\H^V$ and those acting on  $\Gamma(\mathbb{T}^n\!,\!E)$
and the Hilbert space completion of the latter.
Working on the former is an elegant
and convenient way to avoid the bothering work with
local trivializations.

\subsection*{Acknowledgments}

It is a pleasure to thank D. Franco, J. Gracia-Bond\'\i a,
F. Lizzi, R. Marotta, F. Pezzella,
R. Troise, P. Vitale for useful discussions.
We acknowledge support by  the ``Progetto FARO: Algebre di Hopf, differenziali e di
vertice in geometria, topologia e teorie di campo classiche e quantistiche'' of the 
Universit\`a di Napoli Federico II.

\end{document}